\documentclass[useAMS,usegraphicx,usenatbib]{mn2e}
\usepackage{times}
\usepackage{url}

\title[Stokes imaging of AM Her systems \ -- II]{Stokes~imaging~of~AM~Her~systems~using~3D~inhomogeneous models~--~II. Modelling X-ray~and optical data~of~CP~Tucanae.\thanks{Based on observations made at the Observat\'{o}rio do Pico dos Dias, Brazil, operated by the Laborat\'{o}rio Nacional de Astrof\'{i}sica.}}
\author[Silva et al. 2012]
{K.~M.~G. Silva$^{1}$\thanks{E-mail:karleyne@gmail.com},
C.~V. Rodrigues$^{1}$, J.~E.~R. Costa$^{1}$, C.~A. de Souza$^{1}$\thanks{Now at Universidade Federal de Juiz de Fora -- Rua Jos\'{e} LourenÃ§o Kelmer, s/n  -- 36036-330 - Campus Universit\'{a}rio Martelos, Juiz de Fora - MG -- Brazil}, D. Cieslinski$^{1}$
\newauthor
 \& G.~R.  Hickel$^{2}$\thanks{Now at Universidade Federal de Itajub\'{a} -- Av. BPS, 1303 - Pinheirinho - Itajub\'{a} - MG -- Brazil}\\
$^{1}$ Instituto Nacional de Pesquisas Espaciais/MCT --
Av. dos Astronautas, 1758 -- 12227-010 - S\~ao Jos\'e dos Campos - SP -- Brazil\\
$^{2}$ Universidade do Vale do Para\'{i}ba -- Av. Shishima Hifumi, 2911 - Urbanova - S\~ao Jos\'e dos Campos - SP -- Brazil.}
\begin{document}

\date{\today}


\maketitle


\begin{abstract}

The viewing geometry of the polar CP~Tuc that better explains its optical and X-ray light curves is controversial. Previous modelling of white-light polarimetric data considered the partial self-eclipse of an extended inhomogeneous emitting region. Alternatively, phase-dependent absorption has been used to reproduce the X-ray data. This paper presents new optical polarimetric data of CP~Tuc and a model that consistently explains its optical and X-ray data. The model was based on an extension of the {\sc cyclops} code that added X-ray bremsstrahlung emission and pre-shock region absorption to the original version, which only accounted for cyclotron emission. The new code creates the possibility of simultaneous optical and X-ray fitting.  We show that self-eclipse and absorption data have distinct signatures on the X-ray spectra. Although we were able to reasonably fit the CP~Tuc optical data to cases of absorption and self-eclipse, we were only able to reproduce the X-ray orbital modulation after considering the absorption in the pre-shock region. Specifically, we were unable to reproduce the X-ray observations in the self-eclipse case. We found that the primary emitting region in CP~Tuc is located near the rotation pole that approximately points to the observer. 
\end{abstract}


\section[Introduction]{Introduction}
\label{intro}

Cataclysmic variables (CVs) are compact binaries in which a low-mass late type main-sequence star transfers matter to a white dwarf (WD) due to Roche lobe overflow. Magnetic CVs (mCVs) are systems in which the WD possesses a strong magnetic field that controls the dynamics of the accretion flow near the WD. mCVs are divided into two subclasses: AM Her systems (i.e., polars) \citep{cropper1990} and DQ Her systems (i.e., intermediate polars) \citep{patterson1994}. The typical magnetic field on the WD surface of polars and intermediate polars ranges from 5-100~MG and 0.1-10~MG, respectively.

The WD spin of polars is synchronised with the orbital period, and the magnetic field prevents the formation of an accretion disc \citep{cropper1990}. The magnetic pressure overcomes the accretion flow pressure at the Alfv\'{e}n radius. In this so-called coupling region, the flow from the secondary star leaves the ballistic trajectory and turns to a magnetically driven stream to the WD surface, thereby forming an accretion column. Near the WD, the material reaches a supersonic velocity, and a shock front is formed. The region between the shock front and the WD surface is called the post-shock region \citep{lamb1979}. As a result of the shock, this region is hot, with temperatures greater than 10 keV. The region cools down via cyclotron and bremsstrahlung emissions, which dominate the optical and X-ray ranges, respectively.

The cyclotron emission from the post-shock region accounts for the orbital variation of the optical flux and the polarisation of polars. Recently, we developed a code to reproduce the optical emission of mCVs: {\sc cyclops} \citep{costa2009}. This code adopts a 3D-representation of the accretion column and WD, which allows us to study the complex physics and geometry of mCVs. This code calculates the cyclotron emission from inhomogeneous post-shock regions and the attenuation from the pre-shock region by considering the Thomson scattering. \citet{costa2009} briefly review previous models.

The X-ray emission of polars is consistent with hot plasma emission \citep{lamb1979}. The Raymond-Smith and Meka models \citep{raymond2009} can reproduce this emission. Initially, the X-ray data of mCVs were modelled using a single temperature hot plasma continuum and Gaussian profiles to represent the few iron emission lines observed in the spectra \citep{ishida1991,ezuka1999,terada2001}. However, the X-ray spectra of certain mCVs cannot be fit with a single bremsstrahlung component, which indicates the presence of multi-temperature plasma in the post-shock region \citep{ezuka1999}. The radiative cooling in the post-shock region generates the temperature structure \citep{wu1995,done1998,cropper1999}. 

Absorption is necessary to reproduce the spectral energy dependence of mCVs \citep[e.g.,][]{mukai2011}. On one hand, interstellar absorption in the 0.1-2~keV region with typical column densities of 1$\times 10^{21}$cm$^{-2}$ is observed in the X-ray mCV spectra \citep[e.g.][]{ezuka1999}. Moreover, an additional absorption variable in the orbital cycle has been observed in certain systems \citep{done1998,cropper2000}. A partial occultation of the emitting region by the neutral material in the pre-shock region causes this absorption. Absorption models adopt an ad-hoc representation of these geometrical effects \citep{done1998,cropper2000}.  

In addition to absorption, the occultation of the emitting region behind the WD limb can produce orbital variations in the emission of mCVs \citep{king1984}. For instance, \citet{allan1998} discussed these two processes in the context of the phase-dependent effects observed in the intermediate polar EX Hya. Both effects are possible in AM~Her systems \citep{watson1989}. The absorption by the accretion column can produce narrow dips due to the material located near the coupling region, broad dips due to the material near the post-shock region, or both \citep{warren1995,sirk1998,tovmassian2000}.

The X-ray fits of polars are usually conducted without considering the optical data, partially because distinct codes are used to model each spectral range. One exception is the polar RX~J2115-5840 model \citep{ramsay2000}. In this case, the cyclotron 2D-emitting region, obtained from the optical data fit using the methodology of \citet{potter1998}, is used to define the location of the X-ray emission. However, the density and temperature profiles adopted when fitting each energy band are not identical. 

CP~Tuc is a polar that presents energy-dependent orbital X-ray modulations in the high state of brightness \citep{misaki1995,misaki1996}. The \textit{ASCA} satellite discovered CP~Tuc in February 1995 \citep{misaki1995}. Two suggestions have been made regarding the nature of this modulation (as for EX Hya). \citet{misaki1996} fit the phase-resolved spectra using a partial-covering absorption model. Alternatively, \citet{ramsay1999} argued that accretion stream absorption was an unusual mechanism to modulate the X-ray light curves of polars and considered a self-eclipse scenario to be more likely. These authors fit the optical polarimetry data using one self-eclipsed region and suggested that a gradually eclipsed, extended inhomogeneous region can explain qualitatively X-ray modulation. 

This paper presents new optical polarimetric data from CP~Tuc and extends {\sc cyclops} to model optical and X-ray data together. This new version of {\sc cyclops} is then used to study the spectral signature of the X-ray modulation via absorption or self-eclipse as well as how these situations can be distinguished using  high-energy observations of polars. Subsequently, {\sc cyclops} was specifically applied to CP~Tuc. This work is organised as follows: Section \ref{modeld} details the modifications conducted on {\sc cyclops}.  Section \ref{gresults} presents a study on the phase-dependent effects in the X-ray spectra of AM Her systems caused by self-eclipse or partial covering by the pre-shock region. Section \ref{aplication} presents the model of the CP~Tuc optical and X-ray observations, which includes new optical polarimetric data. The results are discussed in light of the previous literature. Section \ref{conclusion} presents the conclusions. The preliminary results of this study were presented in \citet{silva2011a} and \citet{rodrigues2011}.

\section[Extending {\sc cyclops} to X-rays]{Extending {\sc cyclops} to X-rays}
\label{modeld}

{\sc cyclops} is a code designed to reproduce the cyclotron emission that originates in the post-shock region of mCVs. A detailed description of {\sc cyclops} is presented in \citet{costa2009}. This section describes the implementation of the X-ray emission in {\sc cyclops}. Initially, we present an overview of the code (Section \ref{cyc}). Subsequently, we outline the implementation of the bremsstrahlung emission (Section \ref{emi}), the absorption that occurs in the accretion column (Section \ref{abs}), and the procedure for simultaneously fitting optical and X-ray data (Section \ref{fit}).

\subsection[{\sc cyclops}]{{\sc cyclops}}
\label{cyc}

{\sc cyclops} builds a 3D grid to represent the entire accretion column from the threading region to the white dwarf surface. The entire column is divided into several volume elements (voxels) according to the chosen spatial resolution. The lines of a centred dipolar magnetic field define the geometry of the accretion column. The magnetic axis has an arbitrary direction relative to the white dwarf rotation axis. Two footprints (northern and southern) are on the WD surface. This paper considers only one of these emitting regions. We defined the post-shock region from the WD surface to the height of the shock front. The emission comes from this region. The pre-shock region extends from the shock front to the threading region. The phase-dependent absorption is produced in the pre-shock region (see Section \ref{abs}). The functions described below define the electron density and temperature of each voxel of the post-shock region. The magnetic field strength in each voxel was calculated using the magnetic dipole. 

A Cartesian frame was defined for each orbital phase in which the emission was calculated. The z-direction of this frame was the observers line of sight. Therefore, the observer views the accretion column as a 2D set of sight lines. The radiative transfer for each line of sight was solved by considering the emission and absorption processes that were implemented in the code.

The {\sc cyclops} input parameters are as follows:

\noindent
\begin{itemize}
 \item $i$ is the inclination of the orbital plane of the system relative to the observer.
 \item $\beta$ is the angle between the rotation axis and the centre of the northern region.
 \item $f_l$ defines the tangential location of the maxima of the density and temperature. It ranges from 0 to 1. $f_l = 0.5$ stands for a central location.
 \item $\Delta_R$ and $\Delta_{long}$ define the size of the coupling region and, therefore, the footprint of the emitting region on the WD surface. The coupling region is defined as a 2D region that contains a reference point located at a radius $R_{th}$ from the WD surface and at longitude $long_{th}$. $R_{th}$ and $long_{th}$ are defined by $\beta$ and the magnetic field geometry, see description of the geometry of the emitting region in \citet{costa2009}. The coupling region radial length is 1 $\pm$ $\Delta_R$ $R_{th}$, therefore $\Delta_R$ varies from 0 to 1. Its longitudinal extension is longth $\pm$ 2$\Delta_{long}$ (1-$f_l$), $\Delta_{long}$ varies from 0 to 180 degrees.
 \item $h$ is the height of emitting region in WD radius units.
 \item $B_{pole}$ is the intensity of the magnetic field on the magnetic pole.
 \item $B_{lat}$ and $B_{long}$ define the direction of the dipole axis.
 \item  $T_{max}$ and $N_{max}$ are the maxima of the electron temperature and density, respectively.  
\end{itemize}
\noindent

We chose arbitrary functions to describe the variations in temperature and density with regard to height of a specific voxel from the WD surface, $h_{vox}$, which are

\noindent
\begin{equation}
 T(h_{vox})= T_{max} \exp{\left[( 2.5 \left(\frac{h_{vox}}{h}-1\right)\right]} 
 \label{eq_t}
\end{equation}

\noindent and

\noindent
\begin{equation}
 N(h_{vox})= N_{max} \exp{ \left[ -2.5  \sqrt{\frac{h_{vox}}{h}} \right]}.
\label{eq_n}
\end{equation}

These functions resemble the density and temperature structures in the hot and magnetised post-shock regions obtained by \citet{cropper1999} (see their Figure 1) and \cite{saxton2007} (see their Figure 2). However, Equations \ref{eq_t} and \ref{eq_n} do not represent shock solutions and are not necessarily consistent with other model parameters.

With regard to the density and temperature profiles perpendicular to the radial direction, the code allows us to consider a constant profile or modify $T(h_{vox})$ and $N(h_{vox})$ using the following expressions:

\noindent
\begin{equation}
T(d)= T(h_{vox}) \ e^{-\sqrt{d}} 
\end{equation}

\noindent and

\noindent
\begin{equation}
N(d)= N(h_{vox}) \ e^{-\sqrt{d}},
\end{equation}

\noindent where $d$ is the distance from the voxel to the reference point in which the temperature and density values are greatest. These expressions are plausible representations of the tangential decay of temperature and density in the accretion column. They are useful for testing whether the tangential variation of density and temperature affects the mCV emissions.

The first version of {\sc cyclops} only considered the cyclotron emission and bremsstrahlung absorption processes; moreover, it was used to reproduce the optical and infrared polarisation of V834~Cen \citep{costa2009}. 

\subsection{Bremsstrahlung emission in {\sc cyclops}}
\label{emi}

The bremsstrahlung emissivity of a fully ionised magnetised hydrogen plasma ($N_e=N_{ion}$) was calculated for each voxel (k) according to \citet{gronenschild1978}:

\begin{equation}
\label{emissividade}
j_{E,k} = 1.032~\times~10^{-11}~ g ~ N_{e}^{2} ~ E^{-1} ~T ^{-0.5}~e^{-1.16 \times~10 ^{-7}\frac{ E}{T}} ,
\end{equation}

\noindent
where $N_e$ ($cm^{-3}$) is the electron number density, $E$ is energy, $T(K)$ is temperature, and $\textit{g}$ is the non-relativistic Gaunt factor \citep{mewe1986}. 

The emission of a given orbital phase is the sum of the fluxes from all optical paths (see Section \ref{cyc}). The radiative transport is calculated for each optical path from the farthest to nearest voxels by considering the incident radiation from the previous voxels. Physical quantities vary along the optical path following the equations for density and temperature presented above. This variation naturally incorporates the shock structure and optical depth effects, which are important to correctly calculate the emerging flux (see \citealt{cropper1999}). These opacity effects in X-ray wavelengths are important when the column density is greater than $1.3\ 10^{26}$~cm$^{-2}$. The X-ray models use at least 64 voxels in the radial direction to avoid under-sampling the temperature and density functions.  

Considering the thermal case, which is appropriated to mCVs, the specific intensity for a given energy $E$ from each line of sight $l$ at orbital phase $ph$ is
 
\begin{equation}
I_{E,l,ph} = \sum_{k=1}^{nk} (I_{E,l,ph,(k-1)} ~-~bb_{E,ph,k})~e^{-\tau_k}~ + ~ bb_{E,ph,k}~ ,
\end{equation}

\noindent where $bb$ is the Plank function and $nk$ is the number of voxels $k$ at line of sight $l$. $\tau_k$ is the optical depth of a specific voxel, which is calculated as

\begin{equation}
\tau_k=\frac{j_{E,k}}{bb_{E,k}} ~\ s ,
\end{equation}

\noindent where $s$ is the optical path equal to the voxel length.
$I_{E,l,ph}$ can be used to construct the image of the post-shock region in a given orbital phase using the frequency selected. In the absence of absorption, the total observed flux at the orbital phase $ph$ is

\begin{equation}
F_{E,ph} = C * \sum_{l} I_{E,l,ph}  \propto \frac{A_{e}}{D^2} * \sum_{l} I_{E,l,ph} ,
\label{eq_flux}
\end{equation}

\noindent where $C$ is a constant, $A_{e}$ is the emitting area, and $D$ is the distance to the system. 

\subsection[Absorption]{Absorption}
\label{abs}

{\sc cyclops} can account for two sources of absorption: from the interstellar medium (ISM) and from the pre-shock region.

To calculate the photo-absorption cross section for both situations, we used the IDL routine \textit{bamabs} \citep{kashyap2000}. This absorption cross section is a function of temperature and abundances. We considered a homogeneous material with solar abundance and temperature equal to 10\,000~K.

The interstellar absorption is constant along the orbital cycle and depends on the quantity of the interstellar material in the direction of the source. The interstellar column density, $N_{col}^{ISM}$ (cm$^{-2}$), was estimated using FTOOL $N_{H}$ \footnote{\url{http://heasarc.gsfc.nasa.gov/cgi-bin/Tools/w3nh/w3nh.pl}}. $N_{col}^{ISM}$ was used as the upper limit of the interstellar column density of hydrogen in the fit.

The pre-shock region followed the magnetic field lines from the post-shock to the threading regions. All the voxels in the pre-shock region were assumed to have the same density and an upper limit set by the minimum density of the post-shock region (see below). The temperature was also constant.

The pre-shock material absorption is phase dependent. This absorption was considered for the phases in which the pre-shock region fell between the observer and the post-shock region. {\sc cyclops} included the absorption for each line of sight $l$, which were summed to produce the total flux coming from the object in a given orbital phase $i$. In particular, each line of sight has an absorption rate that is proportional to the column density of the pre-shock region.

Considering the absorption from the ISM and the upper portion of the column, the flux observed in each orbital phase $i$ is

\noindent
\begin{equation}
F_{E,ph}= \sum_{l}  F_{E,l,ph} \  e^{-( \ A \ \tau_{pre} + B \ \tau_{MI})},
\end{equation}

\noindent where $\tau_{pre} = \ k_{l} \ s \ \sigma_{ph} \ N_{e}^{min}$,  $\tau_{MI}= \sigma_{ph} \ N_{col}^{ISM}$,  $\sigma_{ph}$ is the photo-absorption cross section, $N_{e}^{min}$ is the minimum number density in the post-shock region, and $k_{l}$ is the number of voxels in the pre-shock region for each $l$. The constants $A$ and $B$ are values between 0 and 1 that defines the fractions of $N_{e}^{min}$ and $N_{col}^{ISM}$, which are used to fit the data. 

\subsection[Fitting data]{Fitting data}
\label{fit}

Our aim is to simultaneously fit the X-ray and optical data of mCVs. To determine the model parameters that best fit the observational dataset, we used the least value of the figure of merit, $\chi^2$, which was defined as

\begin{equation}
 \chi^2= \chi^2_{opt} + add \  \chi^2_{rx}, 
\end{equation}

\noindent where $\chi^2_{opt}$ measures the agreement between a given parameter set and the optical data, and $\chi^2_{rx}$ plays the same role for high-energy data. We used light and polarisation curves in the optical regime, and we used spectra for the X-ray energies. The factor $add$ allowed us to choose the relative contributions of the optical or X-ray data in the fitting procedure. This flexibility is important for achieving a satisfactory fit in both energy ranges.

The code automatically defined the wavelengths and the $ph$ orbital phases for which the model should be calculated using the input datasets. Usually, the orbital resolution of optical data is superior than that of the X-ray spectra. Consequently, a X-ray spectrum was produced for all $ph$ orbital phases of the optical dataset. These spectra were summed over specific orbital phase intervals to obtain the spectra in the $j$ interval phases that correspond to the observed X-ray spectra.

The figures of merit slightly differ in the two spectral domains. Specifically, we calculated the functions $\chi^2_{E_{opt}}$ and $\chi^2_{E_{rx}}$ using:

\begin{equation}
\chi^2_{E_{opt}}=  \sum_{E_{opt}}  \sum_{ph}[d_{E_{opt},ph} \ - \ ( f_{cyc} \ F_{E_{opt},ph} + f^{np}_{E_{opt}})]^2,
\label{eq_fluxo_optico}
\end{equation}

\noindent and 

\begin{equation}
\chi^2_{E_{rx}}=  \sum_{E_{rx}} \sum_{j}[d_{E_{rx},j} \ - \ ( f_{cyc} \ F_{E_{rx},j}^{conv})]^2,
\end{equation}

\noindent where $E_{opt}$ and $E_{rx}$ are the energy in optical and X-rays wavelengths, respectively, $F_{E_{rx},j}$ and $F_{E_{opt},ph}$ are the fluxes calculated using the core of the code, and $d_{E_{rx},j}$ and $d_{E_{opt},ph}$ are the observed (total or polarised) fluxes.
The multiplicative value $f_{cyc}$ was used to normalise the model to compare with the data. This procedure was conducted because the distance and the size of the emitting region were not used to correctly scale the model.  We estimated $f_{cyc}$ using the optical data. The additive constant $F^{np}$ represents the non-polarised emission in optical wavelengths. This emission originates from the stellar components of the system. Each frequency has a different $F^{np}$ value because this component is energy dependent. The superscript $conv$ in the X-ray flux is explained below.

The X-ray flux, as calculated by {\sc cyclops}, $F_{E_{rx},ph}$, cannot be directly compared with the data; we must consider the response of the detection system to fit X-ray data. During the data reduction, we obtained two files: the detector redistribution matrix file ($RMF$), which accounts for detector gain and energy resolution, and the ancillary response file ($ARF$), which accounts for the effective area of the telescope/collimator (including vignetting), filter transmission, detector window transmission, detector efficiency and any additional energy dependent effects (see OGIP Calibration Memo CAL/GEN/92-002\footnote{\url{http://heasarc.gsfc.nasa.gov/docs/heasarc/caldb/docs/memos/cal_gen_92_002/cal_gen_92_002.html}}). We used the {\sc idl} routine conv$_{RMF}$  \citep{kashyap2000} to convolve the original flux from {\sc cyclops}, $F_{E_{rx},ph}$, with the response matrices to obtain the flux for observation comparisons, $F_{E_{rx},ph}^{conv}$:

\begin{equation}
 F_{E_{rx},ph}^{conv}= (ARF ~\ F_{E_{rx},ph}) \times RMF .
\end{equation}

In addition, three other internal parameters were automatically optimised during the fitting procedure:

\begin{itemize}
\item an offset in the model orbital phase, $\delta_{phase}$;
\item the fraction of the upper limit of the electron density in the attenuation region, $A$ (see Section \ref{abs});
\item the fraction of the upper limit of the extinction due to ISM, $B$ (see Section \ref{abs}).
\end{itemize}

As in the original code \citep{costa2009}, the function $\chi^2$ was minimised using two algorithms. First, we used the pikaia algorithm \citep{charbonneau1995} with a broad range of parameters. Then, the best solutions were refined using the amoeba algorithm \cite[e.g.,][]{press1992}.

\section[Orbital variation of the X-ray spectra]{Orbital variation of the X-ray spectra}
\label{gresults}

The orbital variation of high-energy spectra is a common feature in polars (see the Introduction). CP~Tuc presents this type of variation and our primary aim was to reproduce its X-ray spectra (see Section \ref{aplication}). However, before presenting our results regarding CP~Tuc modelling, some general results of {\sc cyclops} should be discussed.

The process of X-ray emission in polars is isotropic, and the emitting region is optically thin; therefore, unlike the optical emission, the emitted flux does not change based on the viewing angle. The orbital variations of the X-ray spectra are explained by the temporary obscuration of the post-shock region. This obscuration might be due to the accretion column or the white dwarf. In the first case, the region that intercepts the sight line is cooler than the post-shock region; therefore, it acts as an absorber. This absorption is most likely the origin of the broad dip observed in certain polars. In the second case, the flux is completely blocked (i.e., the flux from the portion of the post-shock region that is located behind the limb of the white dwarf is not observed). The geometry of the mass flux and the orbital inclination completely define both cases. Hence, {\sc cyclops} is a suitable tool for studying these effects because it incorporates a 3D representation of the system.

\begin{figure*}
\begin{center}
\includegraphics[width=0.8\textwidth,trim= 0.0cm 10.2cm 0.0cm 0.0cm, clip]{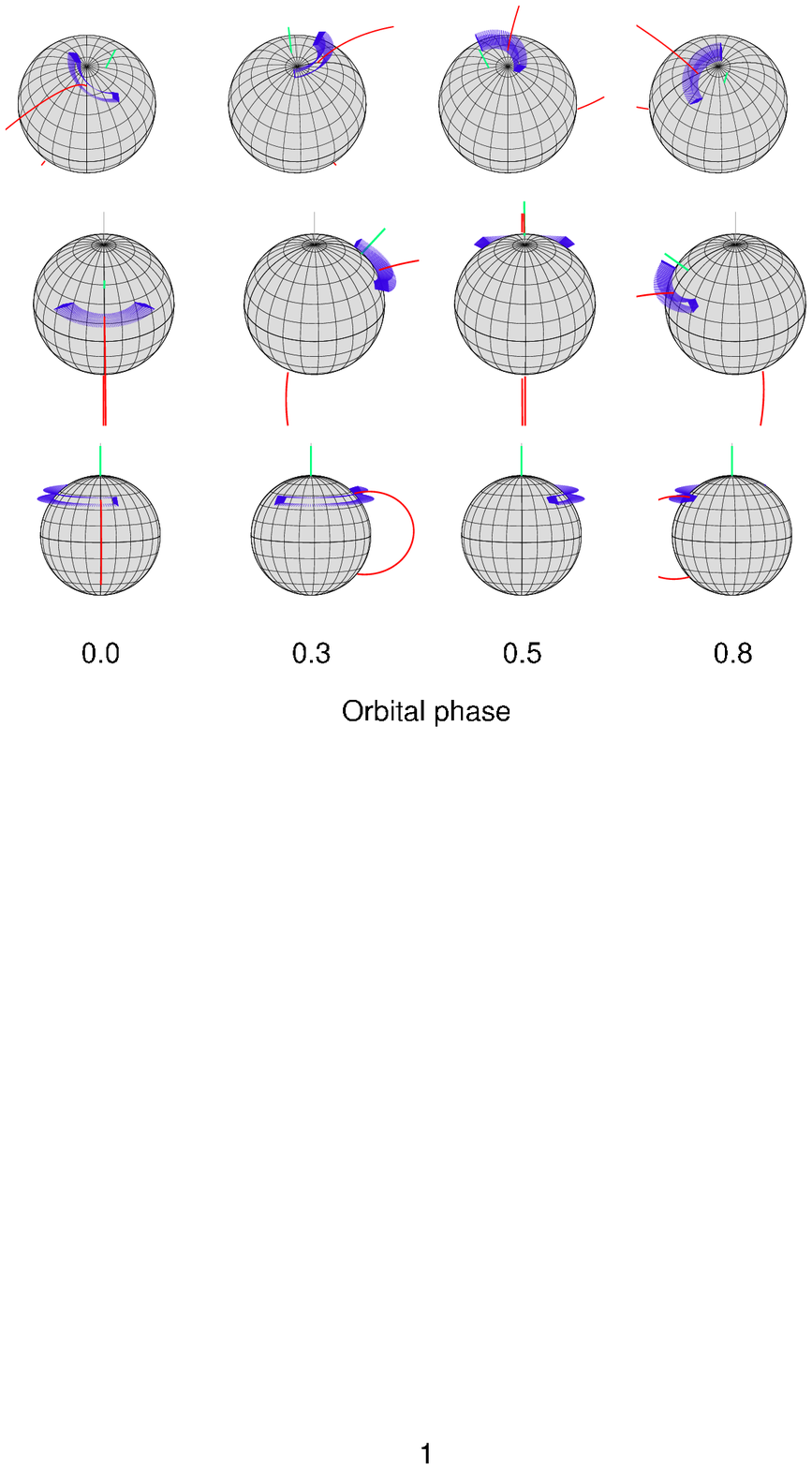}
\end{center}
\vspace{-1.0cm}
\caption{View of the emitting region on the white-dwarf surface for the models \textit{absorption} (top), \textit{self-eclipse I} (middle) and \textit{self-eclipse II} (bottom) - see  Table \ref{model_generic} - in the orbital phases: 0.0, 0.3, 0.5, and 0.8. Only the walls  of the post-shock region (in blue) are represented in this figure. The curved red line is a magnetic field line in the accretion column and the green radial line is the magnetic axis.} 
\label{fig_mod5}
\end{figure*}

\begin{figure*}
\begin{center}
\vspace{-0.8cm}
\includegraphics[width=0.50\textwidth]{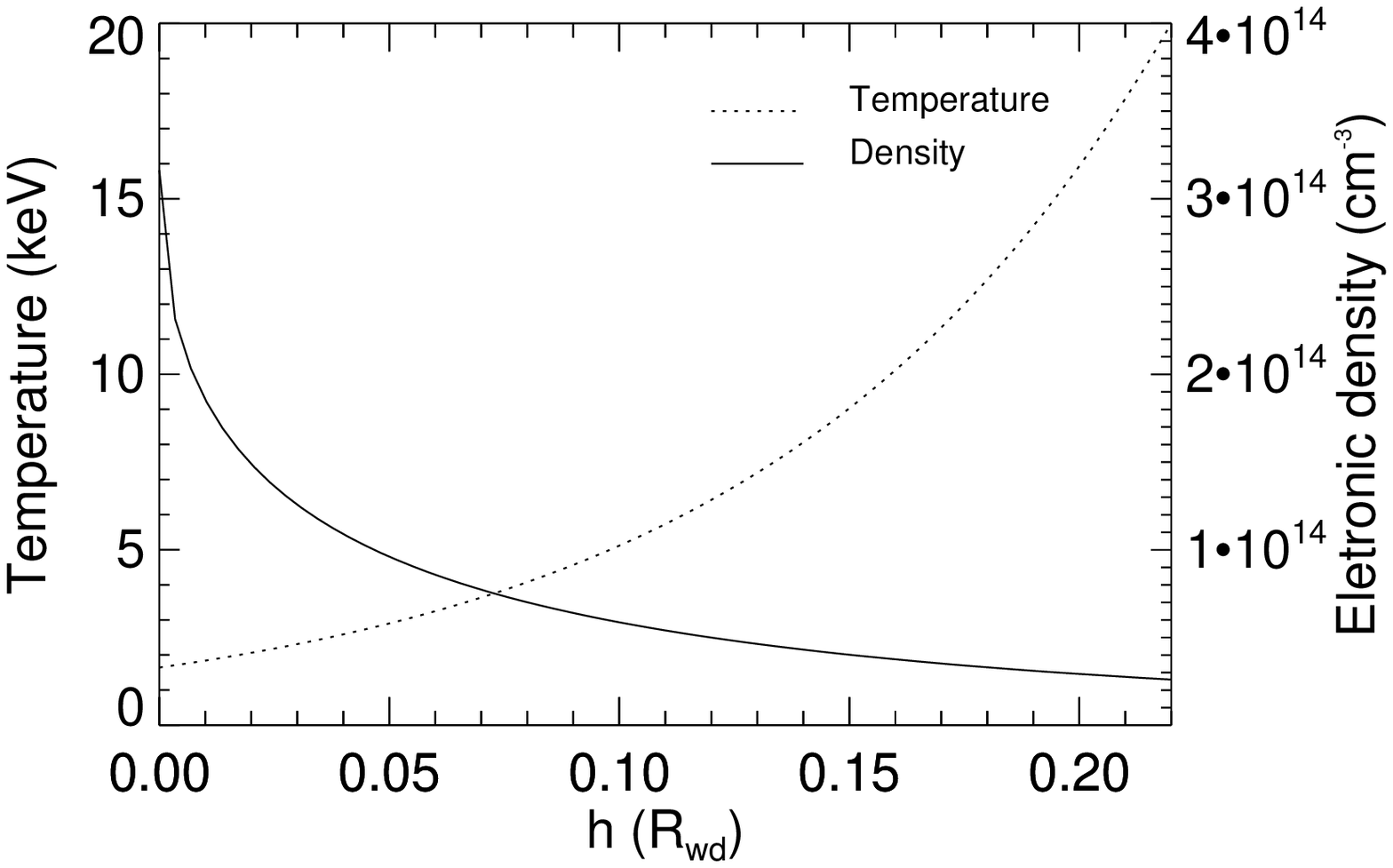}\includegraphics[width=0.50\textwidth]{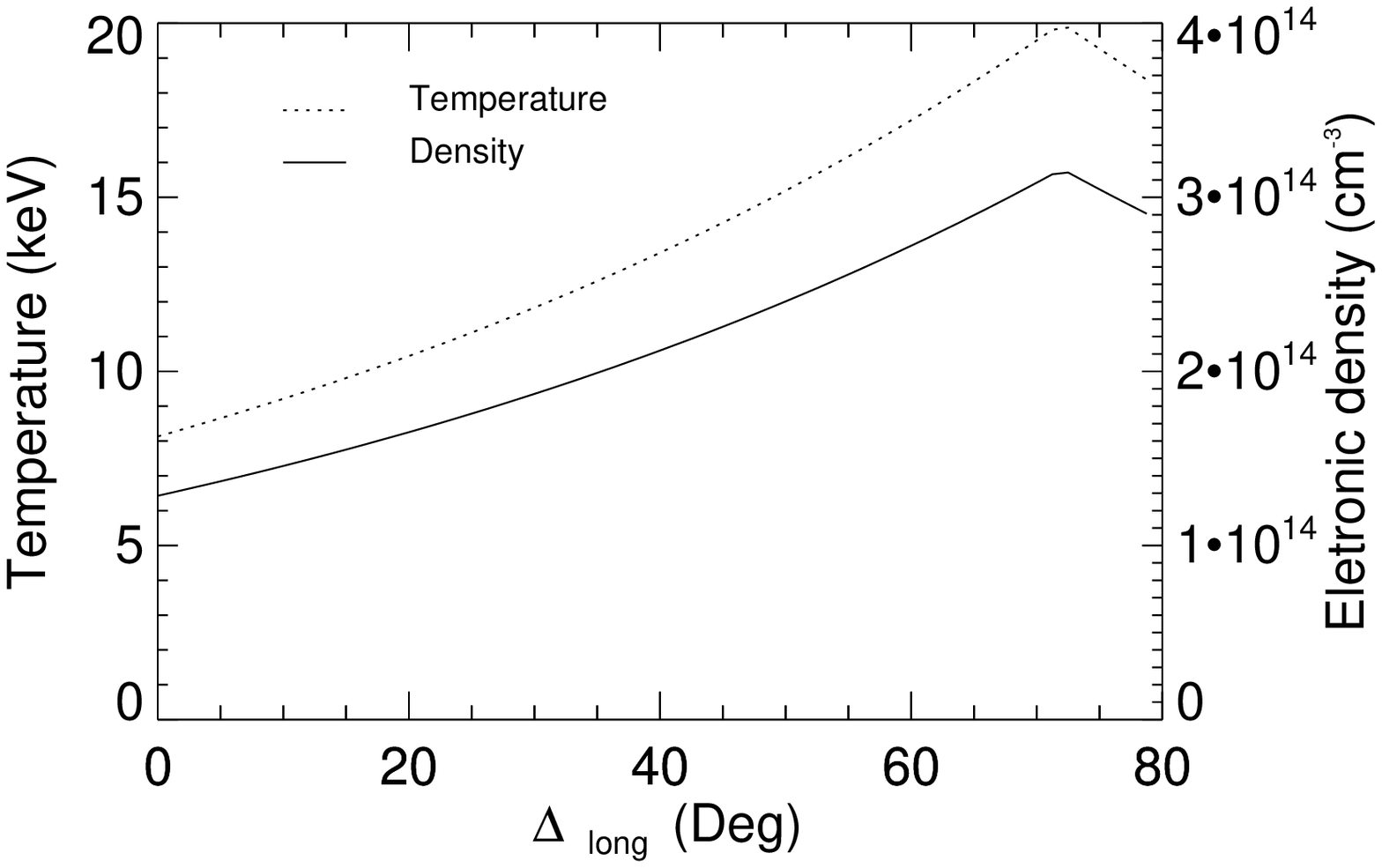}
\end{center}
\vspace{-0.5cm}
\caption{Temperature and density profiles of the models whose parameters are presented in Table \ref{model_generic}. On the left, it is shown the radial profiles used in the three models. On the right, the tangential profiles used in the \textit{self-eclipse II}.}
\label{fig_mod4}
\end{figure*}

We chose three sets of parameters to illustrate the two configurations that can produce spectral orbital variations (see Table \ref{model_generic}). To simplify the discussion, we refer to them as \textit{absorption}, \textit{self-eclipse I}, and \textit{self-eclipse II}. Figure \ref{fig_mod5} shows the system in four orbital phases for each model: 0.0, 0.3, 0.5, and 0.8. Figure \ref{fig_mod4} (left) shows the \textit{radial} variations in density and temperature in the models.  Figure \ref{fig_mod4} (right) shows the \textit{longitudinal} variations in the model \textit{self-eclipse II}. These models are discussed in the following sections.

\begin{table}
\caption{Parameters of the generic models.} 
\begin{center}
\begin{tabular}{l|ccc} \hline
Parameters               & \textit{Absorption}   &  \textit{Self eclipse I} &\textit{ Self eclipse II} \\
\hline
$i$, deg                 & 33           & 58              & 83   \\
$\beta$, deg             & 18           & 68              & 48   \\
$\Delta_{long}, deg$     & 60           & 60              & 80   \\
$\Delta_R$               & 0.20         & 0.20            & 0.10   \\
$h$, $R_{WD}$              & 0.22         & 0.22            & 0.22   \\
$f_l$                    & 0.50         & 0.50            & 0.90    \\
$B_{pole}$, MG           & 20           & 20              & 20   \\
$B_{lat}$, deg           & 74           & 45              & 90   \\
$B_{long}$, deg          & 90           & 360             & 90  \\
$T_{max}$, keV           & 20           & 20              & 20   \\
$N_{max}$, cm$^{-3}$ (log)& 14.5        & 14.5            & 14.5   \\
$A$                      &0.26          & 0               &  0       \\
$B$                      &   0          & 0               &  0       \\
$N_{col} ^1$          &0.21          & --              &  --     \\
$\delta_{phase}$           &0.            & 0.48            & 0.48      \\
\hline
\end{tabular}
\\
$^1$ $N_{col}$ : mean column density of the pre-shock region (10$^{22}$cm$^{-2}$).
\label{model_generic}
\end{center}
\end{table}

\subsection[Absorption]{Absorption}

\begin{figure}
\begin{center}
\vspace{-0.8cm}
\includegraphics[width=0.4\textwidth,trim= 1.5cm 0.cm 1.5cm 0.cm, clip]{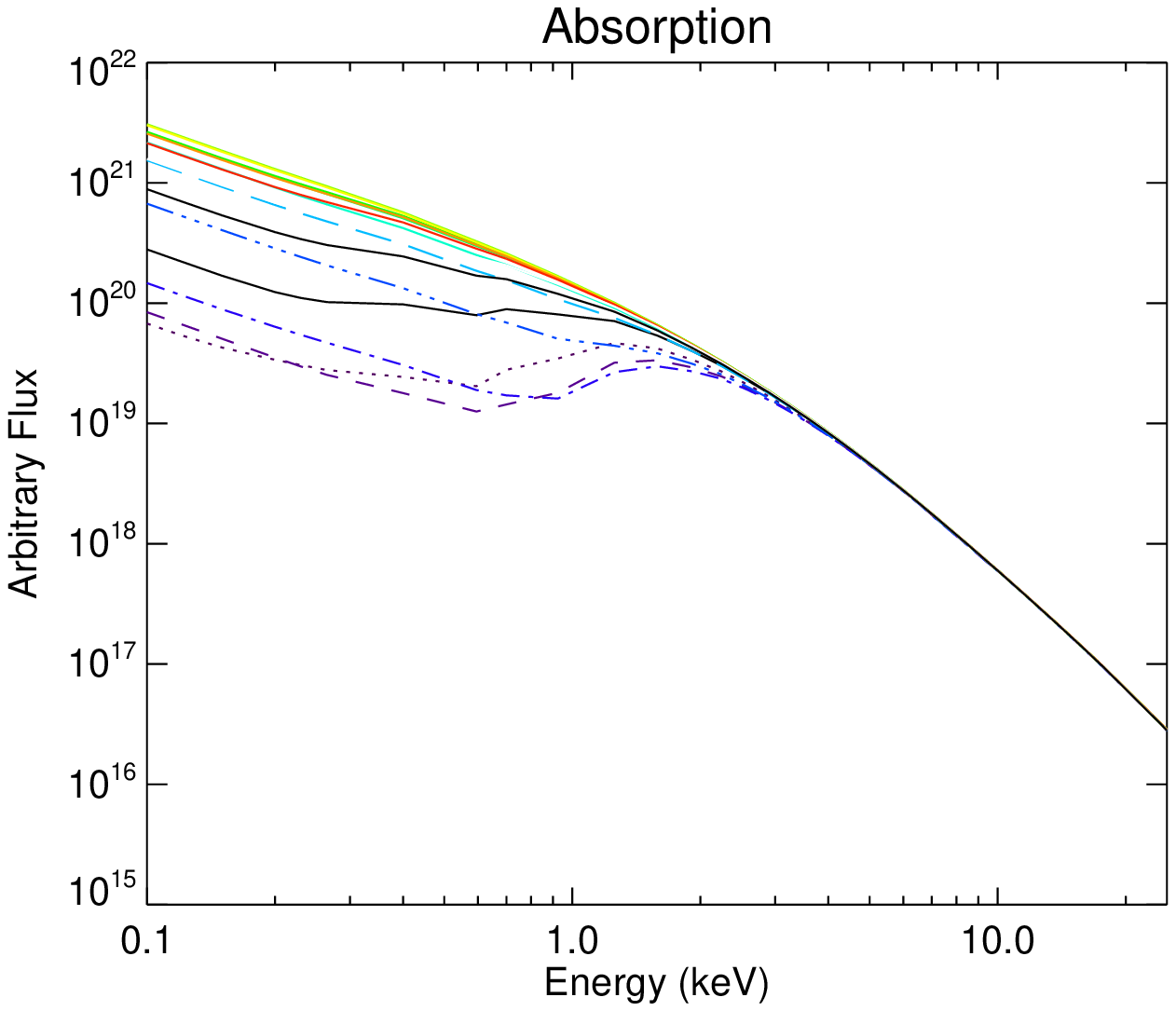}
\includegraphics[width=0.4\textwidth,trim= 1.5cm 0.cm 1.5cm 0.cm, clip]{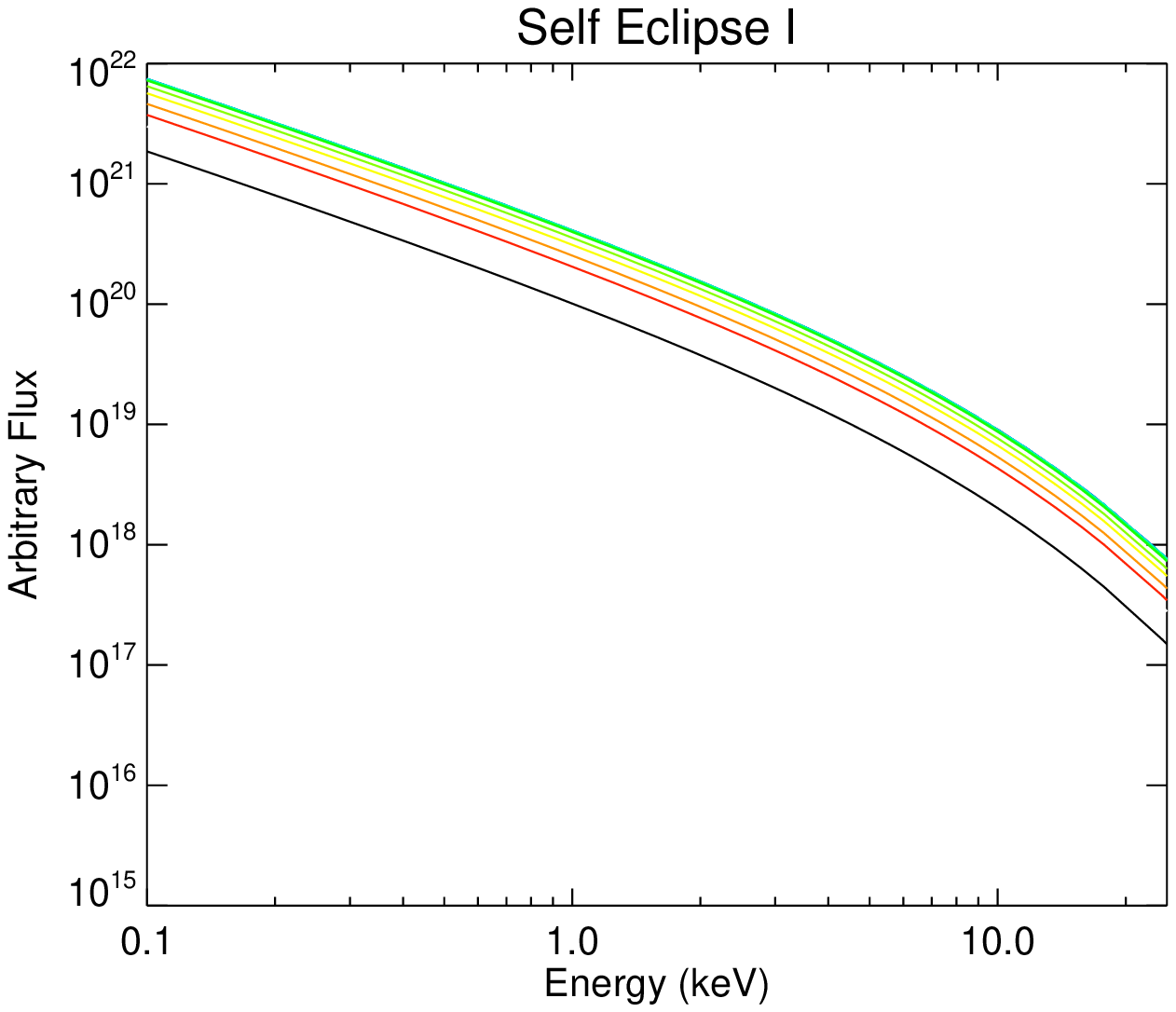}
\includegraphics[width=0.4\textwidth,trim= 1.5cm 0.cm 1.5cm 0.cm, clip]{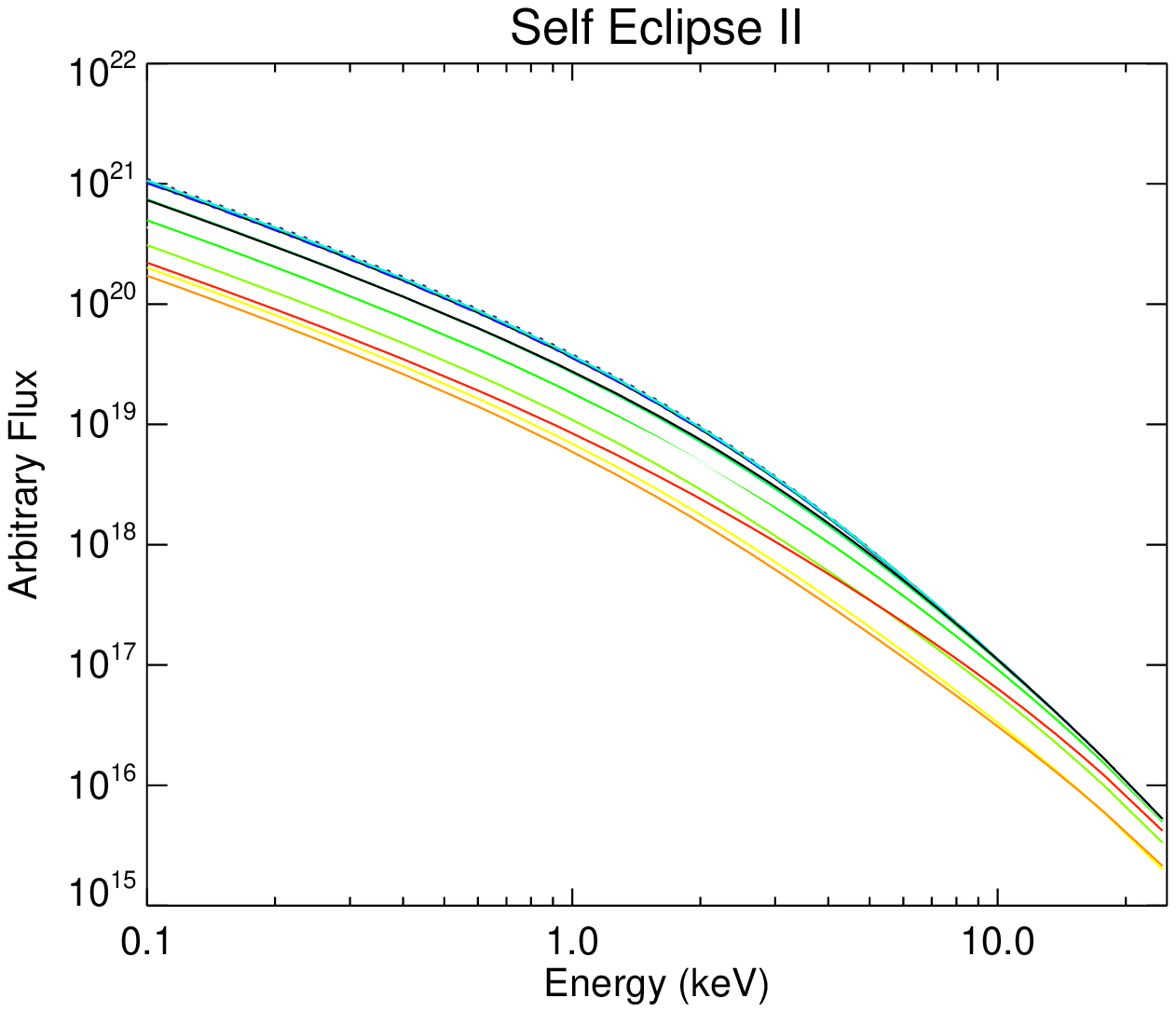}
\end{center}
\vspace{-0.6cm}
\small
\caption{X-ray spectra of the \textit{absorption} (top),  \textit{self-eclipse I} model (middle) and \textit{self-eclipse II} model (bottom) for 15 equally spaced phases. Some phases are overlapped because there is no difference. The spectra are not corrected by the instrumental effects discussed in Section \ref{fit}.}
\normalsize
\label{fig_mod2}
\end{figure}

Figure \ref{fig_mod2} (top) shows the phase-resolved spectra for the \textit{absorption} model (see Table \ref{model_generic} and Figure \ref{fig_mod5}, top panel). The spectra tend to change along the orbital phase due to the variable absorption that is caused by the pre-shock material in the sight line. The absorption varies because (i) the quantity of absorbing material varies, and (ii) the portion of the obscured emitting region varies. The absorption is at its maximum when the region directly points to the observer, and most of the emitting region is observed through the pre-shock region. This scenario occurs during Phase 0 (see Figures \ref{fig_mod5}, left, and Figure \ref{fig_mod2}; magenta short-dashed line). A half-phase later, the region points farther from the observer (top panel, Phase 0.5 in Figure \ref{fig_mod5}), and the pre-shock region is no longer in the observers sight line; consequently, the spectrum does not show absorption (green solid line in the top panel of Figure \ref{fig_mod2}). In addition to this geometrical effect, the photoabsorption cross-section varies as a function of the wavelength. This photoabsorption is larger at lower frequencies, which causes a strong change in the continuum emission with regard to the orbital phase. 
In particular, the spectra did not show difference for energies larger than 3~keV.

\subsection[Self eclipse]{Self eclipse}

In the case of self-eclipses, the occulted fraction of the post-shock region varied along the orbital cycle. This effect was particularly noticeable when the emitting region covered a large area of the WD surface. If the temperature changes along the emitting region, the partial occultation generates spectral changes. We studied two situations that concerned the density and temperature profiles in the post-shock region. 

\begin{itemize}
\item \textit{self-eclipse I} - only radial variation;
\item \textit{self-eclipse II} - only longitudinal variation. 
\end{itemize}

In the case of \textit{self-eclipse I}, the WD partially occulted the region (see Figure \ref{fig_mod5}, middle panel, Phase 0.5). Consequently, the flux decreased for some orbital phases, but significant spectral variation was not found (see Figure \ref{fig_mod2}, middle panel). The emission from the lower layers (i.e., regions closer to the WD surface) might explain the absence of spectral variation, even in the phases when the largest occultation of the post-shock region occurred. These regions most likely dominated the emission because they are denser that the regions closer to the shock front.  

The above result raises the question of whether noticeable spectral variations can be obtained when we consider longitudinal variations in temperature and density. The \textit{self-eclipse II} model (see Table \ref{model_generic} and the lower panel of Figure \ref{fig_mod2}) provides such an example. Using $f_{l}=$ 0.9, the maximum density and temperature of the region are located near one of the extremities of the emitting region and have values twice as large as the other extremity (Figure \ref{fig_mod4}, right).
In this model, the temperature and density decrease outward in a longitudinal direction, whereas density decreases as temperature increases in the radial direction.
During Phase 0.5 (see Figure \ref{fig_mod5}, bottom), the hotter and denser region is occulted, whereas regions with smaller temperatures and densities remain visible. 
Figure \ref{fig_mod2} (lower panel) shows the spectra along all the cycles. The flux levels are different and small changes in the spectral index are also observed. Therefore, longitudinal variations in density and temperature can generate small changes in the shape of spectra. 

\subsection[Discussion]{Discussion}
\label{generic:dis}

The previous sections described an improvement to {\sc cyclops} that models the X-ray emissions of mCVs. The 3D representation of the accretion column provides a physically consistent description of the absorption orbital variation observed in these systems. This new representation contrasts with previous models, which relied on ad-hoc prescriptions of the orbital energy dependence of absorption.

Regarding the X-ray orbital variation observed in polars, we showed that pre-shock absorption primarily affects the low energy region of the spectra, whereas self-eclipse similarly reduces the flux of all energies. This difference can be used to distinguish the origin of orbital variation in polars. In particular, this absorption might better explain the changes in the spectral index of the X-ray spectra of polars. 

\begin{figure}
\begin{center}
\vspace{-0.8cm}
\includegraphics[width=0.50\textwidth,trim= 2cm 0.cm 2.2cm 0.cm,]{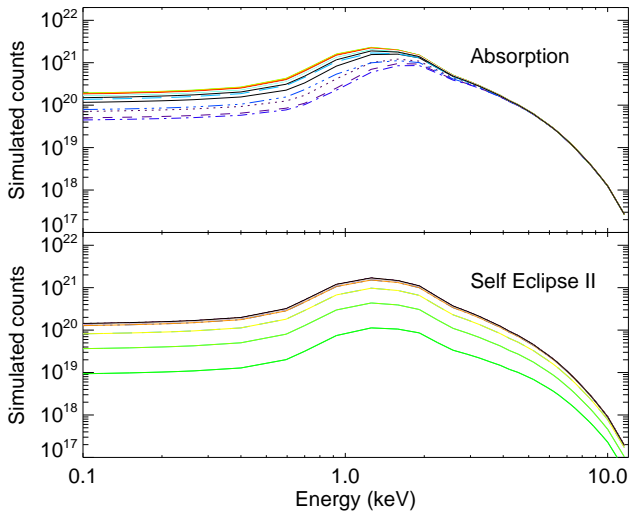}
\end{center}
\vspace{-0.5cm}
\caption{(Up) Spectra in arbitrary units for the \textit{Absorption} model convolved with the ARF and RMF from \textit{ASCA} satellite. (Bottom) The same for the  \textit{self-eclipse II}  case.}
\label{fig_mod6}
\end{figure}

However, the sensitivity of X-ray instruments might conceal the differences described above. To determine whether this hypothesis is true, Figure \ref{fig_mod6} shows the generic models convolved with ASCA GIS response. This instrument has a low response for energies smaller than 1~keV. In this energy range, the differences between the absorption and self-eclipse models are larger, but due to the instrument response these differences are severely attenuated; however, the two models can be distinguished. First, the wavelength of the maximum flux in the absorption model moves (in the 1 -- 2~keV range), whereas this movement is not present in the self-eclipse model. Second, the flux does not change for energies larger than 3~keV in the absorption model.

\section[Application: modelling CP~Tuc]{Application: modelling CP~Tuc}
\label{aplication}

This section describes the simultaneous modelling of the CP~Tuc optical and X-ray data using {\sc cyclops}. Section~\ref{cptuc_intro} presents the previous results concerning CP~Tuc that constrain this modelling. This section also summarises the previous modelling of this polar. 
Section \ref{data} presents the  CP Tuc dataset used in the modelling.
The application of {\sc cyclops} to the CP~Tuc data is presented in Section~\ref{mresults} and is discussed in Section~\ref{discussion}.

\subsection[Introduction on CP~Tuc]{Introduction on CP~Tuc}
\label{cptuc_intro}

CP~Tuc (AX~J2315-592) was discovered by the Japanese satellite $\textit{ASCA}$ \citep{misaki1995}. 
After that report, \citet{thomas1995} found its optical counterpart and spectroscopically confirmed its classification. A deeper analysis of the optical data revealed emission lines that contained the two components usually found in polars: one narrow line associated with the illuminated secondary and one broad line originating from the accretion column near the white dwarf \citep{thomas1996}. The zero point of the spectroscopic ephemeris was defined at the blue-to-red crossing of the narrow component (i.e., the inferior conjunction of the secondary). The broad component presented a maximum blue shift near the 0.9 orbital phase, which indicates that this region has the smallest angle with regard to the line of sight during this phase. \citet{thomas1996} modelled the $I$ band light curve using the cyclotron models of \citet{chanmugam1992} and found a system inclination of 40\degr\ and a magnetic field direction in the emitting region inclined approximately 30\degr\ to the line of sight.  They estimated the cyclotron spectrum of CP~Tuc using the difference between the spectra obtained during bright and faint orbital phases. Using a temperature of 17~keV (as estimated using X-ray data analysis of \citet{misaki1996} - see below), they found that the magnetic field was smaller than 17~MG, and the cyclotron emission was optically thin in optical wavelengths.

\citet{misaki1996} used $\textit{ASCA}$ data to construct light curves at three X-ray energy intervals. The light curve modulations are energy dependent: 87 per cent in 0.7--2.3~keV; 57 per cent in 2.3--6~keV; and 14 per cent in 6--10~keV.  This type of modulation is typical of IPs and usually explained by variable photoabsorption along the orbital cycle. The minimum observed in X-ray coincides with the inferior conjunction of the secondary. \citet{misaki1996} also presented the spectra of CP~Tuc combined in two orbital phase ranges and fit them using a partial covering model. They suggested that the photoabsorption produced by the accretion column modulates the X-ray light curves.

\citet{ramsay1999} obtained additional X-ray observations, optical photometry, and polarimetry for CP~Tuc. The absence of periodicity other than the orbital period and the presence of white light circular polarisation, which reached 10\% and was modulated by the orbital phase, confirmed the polar classification; however, the energy dependent modulation observed in X-ray light curves suggested an IP classification. Significant photometric modulation was not found in the B and V bands, but it reached 1.5 (2.0) mag in the $\textit{R}$ ($\textit{I}$) band. These findings indicate that the system has a weak magnetic field, as previously suggested by \citet{thomas1996}.

Fixing the field at the magnetic pole at 15~MG and the temperature at 17~keV, \citet{ramsay1999} modelled the polarimetric data and found that the inclination, $i$, should be larger than 20\degr. Although they did not find a unique model that best fit the data, they discussed a model with $i$ = 42\degr and $\beta$ = 50\degr. The optical light curve was modelled considering the self-eclipse of an extended region. Moreover, to explain the orbital energy dependence of the X-ray light curves, the authors suggested a temperature distribution along the longitudinal direction of the emitting region. 

\citet{beuermann2007} presented the Zeeman tomography of CP~Tuc. These observations were conducted in June 2000 when the system was in a low state (V=19~mag). Fixing the parameters at those found by \citet{ramsay1999}, \citet{beuermann2007} found two equally possible magnetic field configurations: bipolar or multi-polar. In both situations, the magnetic field is low (approximately 10~MG), and both models reproduce the data equally well.

\subsection[Observational data of CP~Tuc]{Observational data of CP~Tuc}
\label{data}

The optical polarimetric observations and the reduction of CP~Tuc are presented in Section \ref{opt}. Our extraction of the spectra using the X-ray data obtained by \citet{misaki1996} is described in Section \ref{drx}.

\subsubsection[Optical observations]{Optical observations}
\label{opt}

We obtained the CP~Tuc optical data using the Perkin-Elmer 1.6-m telescope at the {\it Observat\'orio do Pico dos Dias} (OPD) operated by the {\it Laborat\'orio Nacional de Astrof\'\i sica} (LNA), using a CCD camera modified by the polarimetric module, as described in \citet{magalhaes1996}. The sensor was an EEV front-illuminated CCD. Table \ref{tabdados} presents a short description of these observations.

\begin{table*}
\begin{center}
\caption{Optical and X-ray data of CP~Tuc.}
\label{tabdados}
\begin{tabular}{l c c c c c }
\hline
Date       & Instrument                           & Filter & $T_{int}$ & Duration    & Ref.   \\
\hline
1997 Aug 29& Polarimeter + CCD106  + $\lambda$/4  & $R_c$ & 90s        &    5~h       &  This work  \\
1997 Aug 30& Polarimeter + CCD106  + $\lambda$/4  & $I_c$ & 90s        &    5.3~h      &   This work \\
1997 Aug 31& Polarimeter + CCD106  + $\lambda$/2  & $I_c$ & 90s        &    5.3~h         &   This work  \\
1995 Nov 24/25& Tek8 CCD                          &  $B$    &  60s       &       -      &  \citet{ramsay1999} \\
1995 Nov 25&  Tek8 CCD                            &  $V$    &  60s       &       -      &  \citet{ramsay1999} \\
1995 Nov 02& \textit{GIS 2,3/ASCA}                &  -    &  -         &    19.13ks  &  \citet{misaki1996} \\
\hline
\end{tabular}
\end{center}
\end{table*}

The reduction followed the standard procedures using {\sc iraf}\footnote{{\sc iraf} is distributed by National Optical Astronomy Observatories, which is operated by the Association of Universities for Research in Astronomy, Inc., under contract with the National Science Foundation} \citep{iraf2,iraf1}. The polarisation was calculated based on \citet{magalhaes1984} and \citet{rodrigues1998} using the package {\sc pccdpack} \citep{pereyra2000} and a set of IRAF routines developed by our group\footnote{\url{http://www.das.inpe.br/~claudia.rodrigues/polarimetria/reducao_pol.html}}. The correction to the equatorial reference system was performed using standard stars. No instrumental polarisation correction was necessary. Our CP~Tuc circular polarisation measurements all had the same sign. However, we were unable to calibrate this sign (whether positive or negative). We adopted the negative sign based on the measurements of \citet{ramsay1999}.

Every linear polarisation measurement has a positive bias: The measured polarisation value is greater than the true polarisation value \citep{simmons1985}. In particular, measurements with $P/\sigma_{P} < 1.4$ provide only the upper limits of the true polarisation value. The polarisation of CP~Tuc was corrected following \citet{vaillancourt2006}.

The ordinary and extraordinary counts of the polarimetric data were summed to obtain light curves. The differential photometry of CP~Tuc was performed using the star USNO~B1.0~0308$-$0806694 (${\rm R_2}$~=~14.4~mag; ${\rm I}$~=~13.57~mag) as a reference. The conversion between ${\rm R_2}$ and Landolt's ${\rm R_C}$ for this object indicated a difference of 0.03 mag \citep{kidger2003}. 

\begin{figure}
\centering
\includegraphics[trim= 2.5cm 1cm 2cm 1.5cm, clip,width=0.55\textwidth]{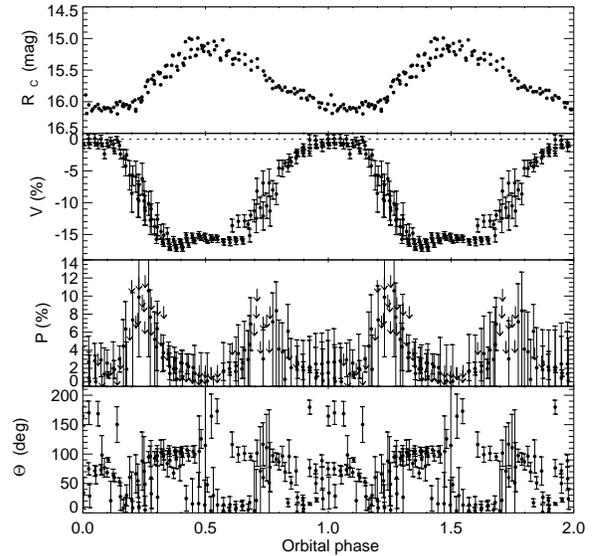}
\caption{Optical light curves and polarisation curves of CP~Tuc in $R_c$ band.  From top to bottom, magnitude, circular polarisation ($V$ in percent), linear polarisation ($P$ in percent), and position angle  ($\theta$ in degrees).} 
\label{fig:lc2}
\end{figure}

\begin{figure}
\centering
\includegraphics[trim= 2.5cm 1cm 2cm 1.5cm, clip,width=0.55\textwidth]{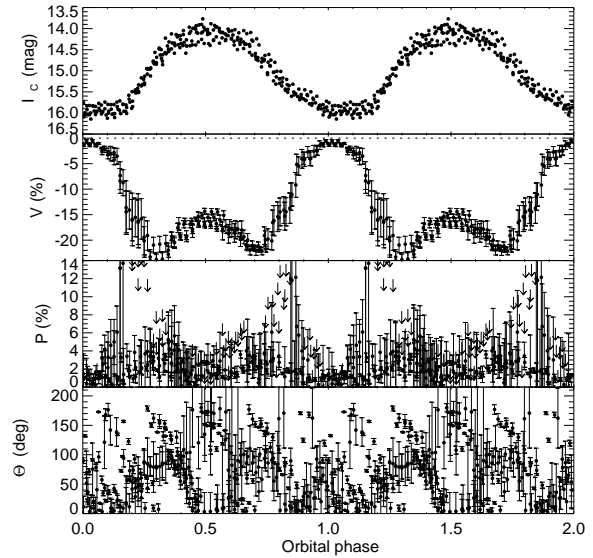}
\caption{Optical light curves and polarisation curves of CP~Tuc in $I_c$ band.  From top to bottom, magnitude, circular polarisation ($V$ in percent), linear polarisation ($P$ in percent), and position angle  ($\theta$ in degrees).} 
\label{fig:lc1}
\end{figure}

Figure \ref{fig:lc2} and \ref{fig:lc1} present the reduced polarimetric data with no binning. From top to bottom, these figures depict magnitude, circular polarisation, linear polarisation, and position angle. 
To phase fold our data we used the CP~Tuc ephemeris obtained by \citet{ramsay1999}. 

For the CP ~Tuc modelling (Section \ref{mresults}), we also considered \citet{ramsay1999}'s $B$ and $V$ photometry obtained at the SAAO 1.0-m telescope. 

\subsubsection[X-ray data]{X-ray data}
\label{drx}

\citet{misaki1996} presented the dataset obtained using the two Gas Imaging Spectrometers (GIS) on board of $ASCA$:  GIS2 and GIS3 (Table \ref{tabdados}). CP~Tuc was outside the field of view for the Solid-state Imaging Spectrometer (SIS) instrument. The response matrices were obtained from the \textit{ASCA} homepage.

\begin{figure}
\centering
\includegraphics[width=0.55\textwidth,trim= 1.5cm 0.7cm 0.cm 1.5cm]{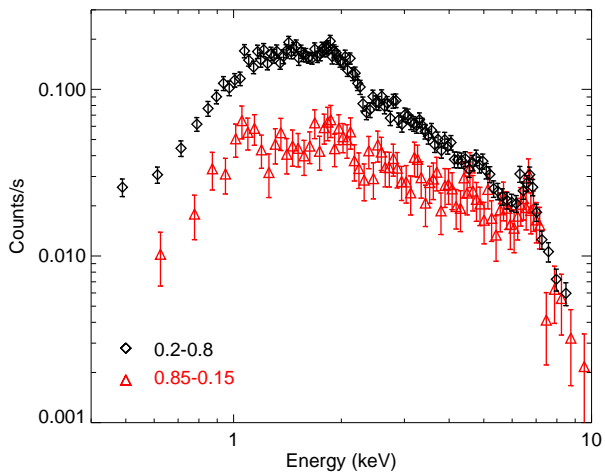}
\caption{Phase-resolved spectra of CP~Tuc obtained using \textit{ASCA} data. Black: the orbital-phase interval of 0.2-0.8. Red: the orbital-phase interval of 0.85-0.15.} 
\label{fig:spec}
\end{figure}

The spectra were extracted at two orbital phase intervals: 0.2-0.8, when the system is brighter, and 0.85-0.15, when the system is fainter. The spectra obtained from GIS2 and GIS3 were combined using the FTOOL \textit{addascaspec} and are shown in Figure \ref{fig:spec}. They do not show significant differences for energies higher than 6~keV, given the errors. A broad iron line is seen at approximately 6.4~keV. 

\subsubsection{CP~Tuc brightness state}

The $R_c$ and $I_c$ light curves of \citealt{ramsay1999} (see their Fig.~1) were obtained during the high state and are consistent with our results (Figures \ref{fig:lc1} and \ref{fig:lc2}). The X-ray observations of Section \ref{drx} were conducted 23 days before \citet{ramsay1999}'s observations and were also obtained during the high state. Therefore, we applied our model to high state X-ray and optical data simultaneously.  

We were unable to use the XMM-Newton observations of CP~Tuc because they were collected during the low state. According to the {\sc AAVSO} observations\footnote{\url{http://www.aavso.org/}}, the system began in the present low state at the end of 2000. In fact, the X-ray observations obtained at the XMM-Newton observatory in 2001 show the system in a low state \citep{ramsay2004}. We analysed the data obtained from the XMM-Newton observatory in 2004 and 2005, when the system was also in a low state. 

\subsection[{\sc cyclops} modelling of CP~Tuc]{{\sc cyclops} modelling of CP~Tuc}
\label{mresults}

As discussed in Section \ref{cptuc_intro}, there are two proposed geometrical scenarios for CP~Tuc: an absorption model \citep{misaki1996} and a self-eclipse model \citep{ramsay2000}. The {\sc cyclops} code handles both configurations (as discussed in Section \ref{gresults}), and it is used here to unravel the geometrical configuration of the magnetic accretion column of CP~Tuc. In a preliminary study, we attempted to model our polarimetric data and found acceptable solutions using any of the CP~Tuc proposed geometries \citep{rodrigues2011}. We called these models Abs1 and SE1, and their parameters are reproduced in Table \ref{cvr_tab}.

In the present modelling, we added the light curves in the \textit{B} and \textit{V} bands from \citet{ramsay2000} as well as two X-ray spectra obtained from \cite{misaki1996}'s data: one in the bright orbital phases (0.2--0.8) and the other in the faint orbital phases (0.85--0.15). The X-ray reduction is described in Section \ref{data}. The region at approximately 6--7~keV, where a strong iron line is located, was not considered in the current fittings. 

We were not interested in fitting the absolute value of flux, so we did not use the distance to CP~Tuc as an input parameter (eq. \ref{eq_flux}). However, we used the same normalisation $f_{cyc}$ across the four optical bands (eq. \ref{eq_fluxo_optico}). This method guaranteed that we adjusted the total and polarised flux spectral dependence in the optical range.

The free parameters of {\sc cyclops} are discussed in Section~\ref{modeld}. 
We adopted  $add$ equals to 10$^3$. Due to the low sensitive of ASCA GIS in low energy, the spectrum is not sensitive to the $B$ value; hence, this parameter was not considered in the CP~Tuc fitting. We used only one frequency to represent each optical band in the search: \textit{B}=6.18$\times10^{14}$~Hz, \textit{V}=5.45$\times10^{14}$~Hz, \textit{$R_{c}$}=4.49$\times10^{14}$~Hz and \textit{$I_{c}$}=3.8$\times10^{14}$~Hz. Using many frequencies to represent a band does not introduce important differences. In fact, the plotted models were calculated considering six frequencies in each band and the filter transmission. The X-ray spectra were calculated using 40 frequencies distributed in the 0.4 to 10 keV interval.

\begin{figure}
\begin{center}
\includegraphics[width=0.6\textwidth,trim= 3cm 1cm 0mm 1.5cm, clip]{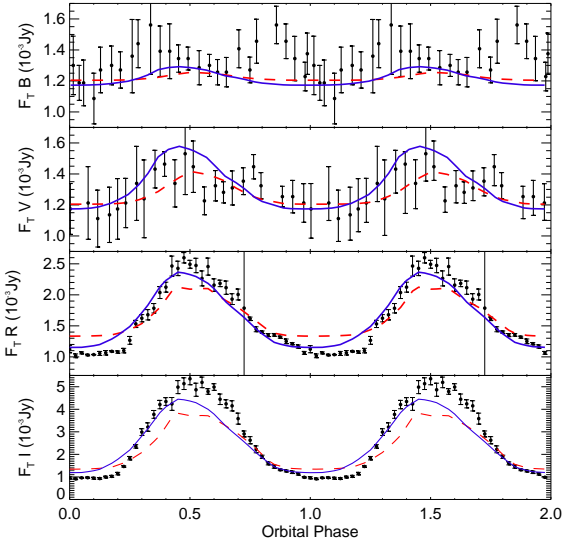}
\end{center}
\caption{Optical light curves of CP~Tuc combined in 40 phase bins. From top to bottom: $\textit{B}$, $\textit{V}$, $\textit{R}_c$ and $\textit{I}_c$ bands. The lines indicate the models Abs2 (red, dashed line) and Abs3 (blue, solid line).The $B$ and $V$ data are from \citet{ramsay1999}.}
\label{fig_flx}
\end{figure}

\begin{figure}
\begin{center}
\includegraphics[width=0.6\textwidth,trim= 3cm 1cm 0mm 1.5cm, clip]{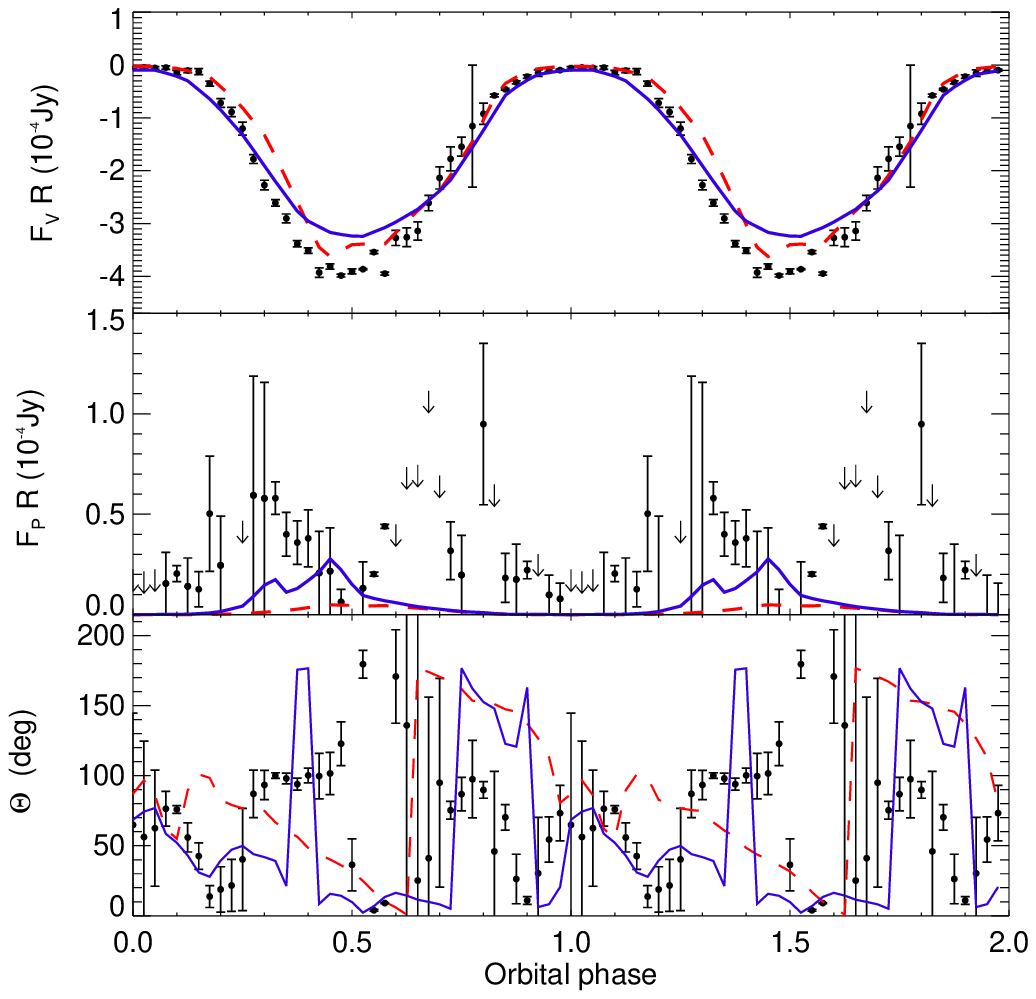}
\includegraphics[width=0.6\textwidth,trim= 3cm 1cm 0mm 1.5cm, clip]{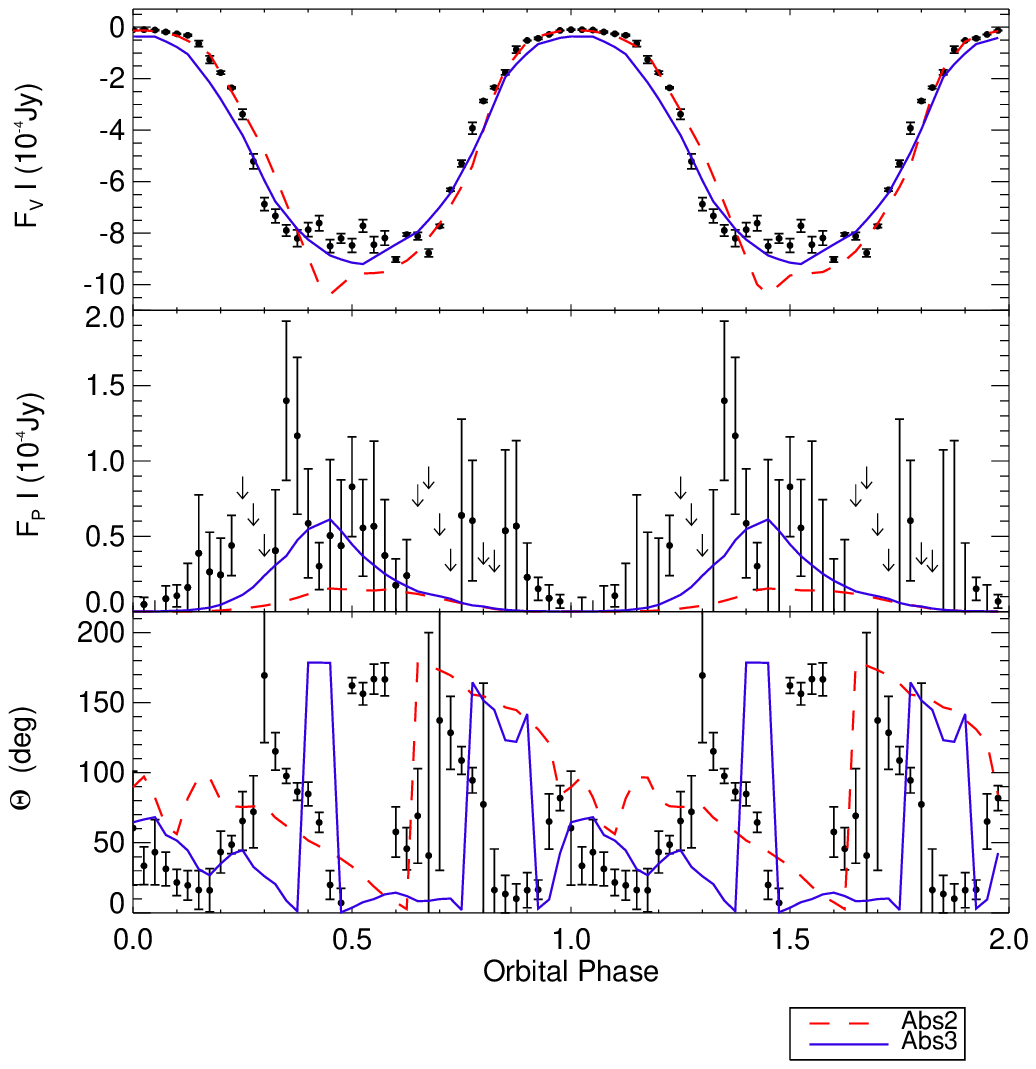}
\end{center}
\caption{Polarimetric curves of CP~Tuc combined in 40 phase bins for $\textit{R}_c$ (top panel) e $\textit{I}_c$ (bottom panel) bands. The lines indicate the models Abs2 (red, dashed line) and Abs3 (blue, solid line). From top to bottom, circular polarised flux ($F_V$), linear polarised flux ($F_P$) and angle of the linear polarisation ($\theta$).}
\label{fig_pol}
\end{figure}

\begin{figure}
\centering
\begin{center}
\includegraphics[width=0.6\textwidth,trim= 1.5cm 0.6cm 0mm 1.5cm, clip]{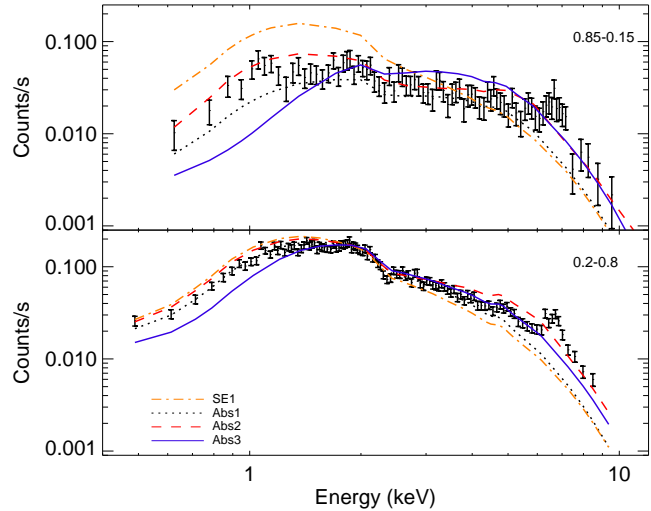}
\end{center}
\caption{X-ray phase resolved spectra in arbitrary units of CP~Tuc for  phase range 0.85-0.15 (top panel) and 0.2-0.8 (lower panel). The lines represent the models.} 
\label{cptuc_abs}
\end{figure}

We determined the best models using two approaches: (i) probing the optical polarisation models from \citet{rodrigues2011}, Abs1 and SE1, and (ii) performing "blind" {\sc pikaia} trials in a larger region of the parameter space. The best fit models are Abs2 and Abs3 (see Figures \ref{fig_flx}, \ref{fig_pol} and \ref{cptuc_abs} and Table \ref{cvr_tab}). The Abs1, Abs2 and Abs3 models have absorption geometries, and the SE1 model presents a self-eclipse of the emitting region. The comparison between Figures \ref{fig_mod6} and \ref{fig:spec} suggested that self-eclipse models were not appropriate for CP~Tuc X-ray data. In fact, the SE1 model did not reproduce the CP~Tuc X-ray spectra (Fig. \ref{cptuc_abs}, dot-dashed line). Nevertheless, we included the self-eclipse domain in the search to verify the robustness of the results of Section \ref{generic:dis}. But no self-eclipse model better than SE1 was found. Thus, we discarded self-eclipse models for CP~Tuc. We henceforth discuss Abs2 and Abs3, which have similar $\chi^2$ values but different geometrical configurations.

\begin{table}
\caption{Parameters of CP~Tuc models.} 
\begin{center}
\begin{tabular}{l|cccc} 
\hline
Parameters               & Abs 1   &  SE 1    & Abs 2       & Abs 3       \\  
\hline
$i$        , deg         & 29      & 22       &  17         &   33         \\
$\beta$, deg             & 24      & 42       &  22         &   36         \\
$\Delta_{long}, deg$     & 18      & 13       &  16         &   32         \\
$\Delta_R$               & 0.32    & 0.52     &  0.12       &   0.65        \\
$h$, $R_{WD}$              & 0.10    & 0.22     &  0.05       &   0.14        \\
$f_l$                    & 0.5     & 0.5      & 0.5         &   0.5         \\
$B_{pole}$, MG           & 6.0     & 6.1      &  6.0        &   7.6         \\
$B_{reg}^1$, MG             & 4.2-5.8 &2.7-5.9   & 4.1-5.3     & 3.8-7.4      \\
$B_{lat}$, deg           & 85      & 35       &  48         &   40         \\
$B_{long}$, deg          & 41      & 336      &  63         &   337         \\
$T_{max}$, keV           & 66      & 102      &  60         &   89         \\
$T_{pond}^2$ , keV        & 11.0   & 18.4       &  10.0       &  15.0        \\   
$N_{max}$, cm$^{-3}$ (log)& 15.7    & 15.0     &  16.4       &   13.8        \\
$A^3$                      &0.26     & 0        &  0.28       &   0.99        \\
N$_{col}^{4}$           &3.9      & --       &  3.0        &   17.0       \\
$\delta_{phase}$         &-0.002   & -0.095   &  0.095      &  -0.12       \\
$\chi^2$ total     & 0.09    & 0.22     &  0.06       &   0.06      \\
$f^{np} (B)$, mJy               & 12  & 11   & 12     &  12  \\  
$f^{np} (V)$, mJy               & 10  & 9   & 11     &  9  \\  
$f^{np} (R_c)$, mJy             & 11  & 11    & 14     &  12 \\  
$f^{np} (I_c)$, mJy             & 14  & 14   & 17     &  13  \\  
\hline 
\end{tabular} 
\end{center}
$^1$ $B_{reg}$: magnetic field range in the post-shock region.

$^2$ $T_{pond}$: mean temperature weighted using the square density, see Section \ref{discussion} for details.

$^3$ $A$: the fraction of the maximum possible electron density in the attenuation region.

$^4$ $N_{col}$: mean column density of the pre-shock region (10$^{22}$cm$^{-2}$).

\label{cvr_tab}

\end{table}

\subsection{Modelling discussion}
\label{discussion}

We begin the discussion of our CP~Tuc models by comparing certain parameters with previous estimates. Zeeman tomography indicated that $B_{pole}$ = 19.8~MG (in the bipolar case) and that the most frequent magnetic field in the photosphere was 10~MG \citep{beuermann2007}. This technique probes the magnetic field on the entire WD surface. We found that $B_{pole}$ = 6~MG and 7.6 MG for Abs2 and Abs3, respectively. Table \ref{cvr_tab} shows the range of magnetic fields in the post-shock region ($B_{reg}$) for each model. For instance, Abs3 has a range of 3.8--7.4~MG. Therefore, the magnetic field found using {\sc cyclops} was smaller that that found by \citet{beuermann2007}. The {\sc cyclops} emitting region reached a height of $0.14~R_{WD}$, which corresponds to a decrease greater than 30\% in the magnetic field relative to the value in the photosphere. Therefore, the radial extension of the emitting region can explain the difference. Zeeman tomography indicated that the CP~Tuc magnetic field configuration is an offset dipole or a multipole \citep{beuermann2007}. As the {\sc cyclops} emitting region covers a small fraction of the WD surface, our technique is not sensitive to distinct large-scale magnetic field configurations.

\citet{misaki1996} estimated a temperature of 17~keV for the post-shock region. The \textit{maximum} temperature of Abs2 and Abs3 were 60 and 89~keV, respectively. However, the mean temperature weighted by the square density, $T_{pond}$, is a more proper comparison with \citet{misaki1996}'s single-temperature model. It is 10~keV for Abs2 and 15~keV for Abs3. Thus, although we used $T_{max}$ to define the temperature profile, the hottest regions did not dominate the X-ray emission. Given that all the gravitational energy is converted to X-ray radiation in the post-shock region, we estimated the lower limit for the WD mass in CP~Tuc using the WD mass-radius relationship from \citet{nauenberg1972}. We found 0.38 and 0.47~$M_\odot$ for Abs2 and Abs3, respectively. The secondary mass was 0.08~$M_\odot$ from the orbital period versus the secondary mass relation (see \citet{knigge2011}). The resulting mass ratios were 0.21 and 0.17 for Abs2 and Abs3, respectively.

\begin{figure*}
\begin{center}
\includegraphics[width=1.0\textwidth,trim= 0.0cm 14.1cm 0.0cm 0.0cm, clip]{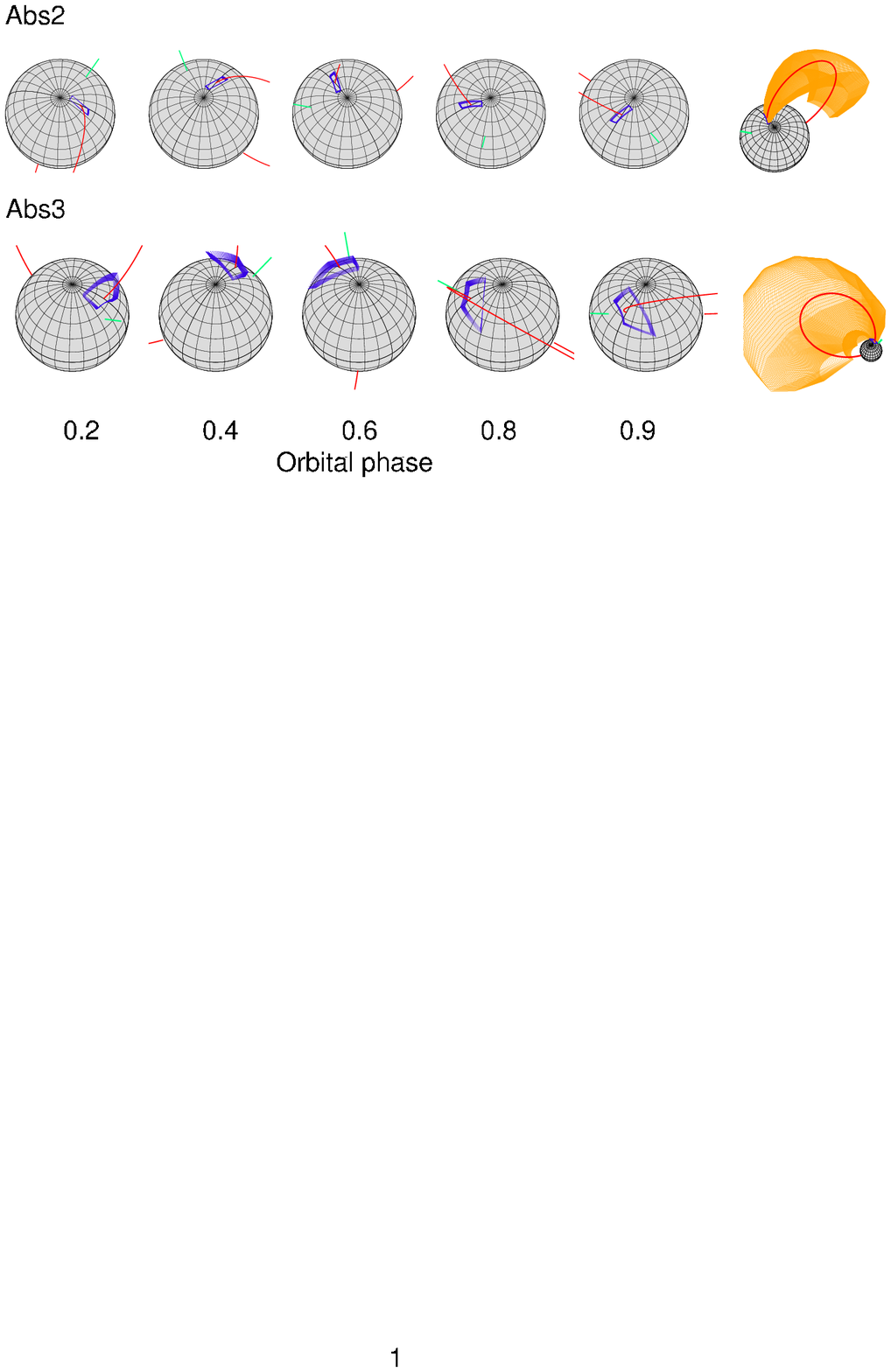}
\end{center}
\caption{Left: View of the emitting region of the models Abs2 (top) and Abs3 (bottom) (Table \ref{model_generic}) in five orbital phases: 0.2, 0.4, 0.6 and 0.8 and 0.9. Only the laterals (blue) of the post-shock region are represented in this figure. Right: View of the accretion column (orange) for models Abs2 in phase 0.6 (top) and Abs3 in phase 0.4 (bottom). In all figures, the curved red line near is a magnetic field line in the accretion column and the green radial line is the magnetic axis. }
\label{fig_res1}
\end{figure*}

Optical spectroscopy fo CP~Tuc \citep{thomas1996} is one observational constraint that can help to distinguish Abs2 from Abs3. The radial velocity of the broad component of the optical emission lines displays sinusoidal behaviour with a maximum redshift during Phase 0.9. Hence, the base of the accretion column must be observed from the smallest angle at Phase 0.9. Figure \ref{fig_res1} shows the system configuration for Abs2 and Abs3 for Phases 0.2, 0.4, 0.6, 0.8 and 0.9. The curved red line threaded through the emitting region is a magnetic field line, which aids the visualisation of the flow direction in the emitting region. The right panels show the entire accretion column and illustrate the pre-shock region geometry. The system configuration of Abs3 has matched the spectroscopy. We observe the region from the top in Phase 0.9. Thus, the post-shock emission has a maximum redshift. The region is pointing away from the observer and has the maximum blueshift at approximately Phase 0.4. In Abs2, the field line points to the observer at approximately Phase 0.2 (maximum redshift) and is in the opposite direction during Phase 0.8 (maximum blueshift), which has not matched the spectroscopy. 

Despite the better agreement between Abs3 and the spectroscopy, we should take this result with caution because it is qualitative. Moreover, the model did not account for certain effects that might improve the fit. The Abs2 model presents the best fit of the X-ray spectra (see Figure  \ref{cptuc_abs}, long-dashed line) and an acceptable fit of the optical data (see Figures \ref{fig_flx} and \ref{fig_pol}, long-dashed line). Its linear polarisation and cyclotron flux modulation were lower than observed. A better agreement between the optical data and Abs2 might be achieved by including an ellipsoidal variation component in the model, because this emission has the proper phasing and colour. The elipsoidal variation was not accounted for in the model.

The Abs3 model (solid lines in Figures \ref{fig_flx}, \ref{fig_pol} and \ref{cptuc_abs}) provides the best fit to the optical data, with linear polarisation and flux modulation consistent with the observations. It also presented an acceptable fit to the high-energy X-ray spectrum, but it overestimated the absorption at low energies. This excess absorption might be due to the adopted approximations of the pre-shock material, which was cold, homogeneous, and had solar abundance. A warmer absorber with smaller abundance might have less photoabsorption.  A decrement in the density of regions farther from the WD might also account for the smaller net absorption.

Importantly, {\sc cyclops} only includes the bremsstrahlung continuum emission in the X-ray range, which is the dominant process. However, other processes might also contribute including Compton scattering in the post-shock region \citep{suleimanov2008}, heating of the WD surface, and emission lines. 

Taking into account all of the above considerations, small differences between the observed data and the model were expected. On the other hand, the quantity and quality of the dataset might be not enough to justify the inclusion of these "second-order" processes in the models. However, we argue that these data are enough to restrict the correct viewing aspect of this system, which has phase-resolved absorption in X-rays.

\section[Conclusions]{Conclusions}
\label{conclusion}

We extended the {\sc cyclops} code to study the X-ray and optical continua emission of polars. We included bremsstrahlung X-ray emission, the dominant radiative process in the post-shock region, and photo-absorption by the pre-shock accretion column. Cyclotron emission and self-eclipse by the white dwarf were already available in the previous version of the code. The new version allowed us to simultaneously model X-ray and optical data.

The calculated spectra accounted for the temperature and density structure in the post-shock region in both radial and tangential directions. Although the X-ray emission of polars is usually optically thin, our radiative transfer solution adequately treated optically thick and thin regions. The magnetic field lines in the 3D space defined the accretion column geometry. This approach resulted in a consistent representation of the geometrical effects of the white dwarf occultation and the pre-shock region absorption. Moreover, the code provided phase-resolved X-ray spectra.

We showed that phase-resolved X-ray spectra have distinct signatures in cases of self-eclipse and absorption. Absorption causes a clear variation of the spectral index along the orbital cycle, whereas these differences are barely distinguished in the self-eclipse case, even when we consider the inhomogeneous distribution of the temperature and density in the emitting region.

We chose CP~Tuc to explore the differences between self-eclipse and absorption in the X-ray spectra of polars because it is a polar with two diverging explanations for its X-ray data, and each one is consistent with one of the above scenarios. We modelled CP~Tuc using the X-ray and optical data from the literature as well as new optical $\textit{R}_{\rm C}$ and $\textit{I}_{\rm C}$ polarimetry and photometry data. 

Our CP~Tuc models indicated a single accretion region object in which cyclotron beaming generated the optical modulation and photo-absorption generated the X-ray modulation, in agreement with \citet{misaki1996}. We could not create a model that described the X-ray spectra using the self-eclipse scenario. The magnetic field intensity of our best fittings was consistent with previous Zeeman tomography estimates \citep{beuermann2007}. The temperature structure matched the single-temperature model of the X-ray continuum \citep{misaki1996}. One of the best fittings had the correct phasing to account for the radial velocity curve of the broad component of the optical emission lines \citep{thomas1996}.

Future improvements to the code are a more physically consistent description of the pre-shock and post-shock density and temperature distributions and inclusion of other emitting process. With regard to CP~Tuc, a more detailed model requires more quality data in high-energy range. 

\section*{Acknowledgements}

This research used: the USNOFS Image and Catalogue Archive operated by the United States Naval Observatory, Flagstaff Station (http://www.nofs.navy.mil/data/fchpix/); the SIMBAD database, operated at CDS, Strasbourg, France; the NASA's Astrophysics Data System Service; and the High Energy Astrophysics Science Archive Research Center (HEASARC), operated by Astrophysics Science Division at NASA/GSFC and the High Energy Astrophysics Division of the Smithsonian Astrophysics Observatory (SAO). KMGS thanks {\sc fapesp} for their financial support (Proc. 2008/09619-5). KMGS thanks CERTOES and COSPAR for the support and training in X-ray data reduction and analysis. CVR acknowledges  {\sc fapesp} (Proc. 2010/01584-8) and CNPq (Proc. 308005/2009-0). We also acknowledge an anonymous referee for uselful suggestions.


\begin{thebibliography}{}

\bibitem[\protect\citeauthoryear{Allan, Hellier \& Beardmore}{1998}]{allan1998} Allan A., Hellier C., Beardmore A., 1998, MNRAS, 295, 167 
\bibitem[\protect\citeauthoryear{Beuermann et al.}{2007}]{beuermann2007} Beuermann K., Euchner F., Reinsch K., Jordan S., G{\"a}nsicke B.~T., 2007, A\&A, 463, 647 
\bibitem[\protect\citeauthoryear{Chanmugam}{1992}]{chanmugam1992} Chanmugam G., 1992, ARA\&A, 30, 143 
\bibitem[\protect\citeauthoryear{Charbonneau}{1995}]{charbonneau1995} Charbonneau P., 1995, ApJS, 101, 309
\bibitem[\protect\citeauthoryear{Costa \& Rodrigues}{2009}]{costa2009} Costa J.~E.~R., Rodrigues C.~V., 2009, MNRAS, 398, 240
\bibitem[\protect\citeauthoryear{Cropper}{1990}]{cropper1990} Cropper M., 1990, Space Science Reviews, 54, 195 
\bibitem[\protect\citeauthoryear{Cropper et al.}{1999}]{cropper1999} Cropper M., Wu K., Ramsay G., Kocabiyik A., 1999, MNRAS, 306, 684
\bibitem[\protect\citeauthoryear{Cropper, Wu \& Ramsay}{2000}]{cropper2000} Cropper M., Wu K., Ramsay G., 2000, NewAR, 44, 57 
\bibitem[\protect\citeauthoryear{Done \& Magdziarz}{1998}]{done1998} Done C., Magdziarz P., 1998, MNRAS, 298, 737
\bibitem[\protect\citeauthoryear{Ezuka \& Ishida}{1999}]{ezuka1999} Ezuka H., Ishida M., 1999, ApJS, 120, 277 
\bibitem[\protect\citeauthoryear{Gronenschild \& Mewe}{1978}]{gronenschild1978} Gronenschild E.~H.~B.~M., Mewe R., 1978, A\&AS, 32, 283 
\bibitem[\protect\citeauthoryear{Ishida et al.}{1991}]{ishida1991} Ishida M., Silber A., Bradt H.~V., Remillard R.~A., Makishima K., Ohashi T., 1991, ApJ, 367, 270 
\bibitem[\protect\citeauthoryear{Kashyap \& Drake}{2000}]{kashyap2000} Kashyap V., Drake J.~J., 2000, BASI, 28, 475 
\bibitem[\protect\citeauthoryear{Kidger}{2003}]{kidger2003} Kidger M.~R., 2003, A\&A, 408, 767 
\bibitem[\protect\citeauthoryear{King \& Shaviv}{1984}]{king1984} King A.~R., Shaviv G., 1984, MNRAS, 211, 883
\bibitem[\protect\citeauthoryear{Knigge, Baraffe \& Patterson}{2011}]{knigge2011} Knigge C., Baraffe I., Patterson J., 2011, ApJS, 194, 28 
\bibitem[\protect\citeauthoryear{Lamb \& Masters}{1979}]{lamb1979} Lamb D.~Q., Masters A.~R., 1979, ApJ, 234, L117 
\bibitem[\protect\citeauthoryear{Magalh\~aes, Benedetti \& Roland}{1984}]{magalhaes1984} Magalh\~aes A.~M., Benedetti E., Roland E.~H., 1984, PASP, 96, 383 
\bibitem[\protect\citeauthoryear{Magalh\~{a}es et al.}{1996}]{magalhaes1996} Magalh\~{a}es A.~M., Rodrigues C.~V., Margoniner V.~E., Pereyra A., Heathcote S., 1996, ASPC, 97, 118
\bibitem[\protect\citeauthoryear{Mewe, Lemen \& van den Oord}{1986}]{mewe1986} Mewe R., Lemen J.~R., van den Oord G.~H.~J., 1986, A\&AS, 65, 511
\bibitem[\protect\citeauthoryear{Misaki et al.}{1995}]{misaki1995} Misaki K., et al., 1995, IAUC, 6260, 1 
\bibitem[\protect\citeauthoryear{Misaki et al.}{1996}]{misaki1996} Misaki K., Terashima Y., Kamata Y., Ishida M., Kunieda H., Tawara Y., 1996, 
ApJ, 470, L53 
\bibitem[\protect\citeauthoryear{Mukai}{2011}]{mukai2011} Mukai K., 2011, xru..conf, 11]
\bibitem[\protect\citeauthoryear{Nauenberg}{1972}]{nauenberg1972} Nauenberg M., 1972, ApJ, 175, 417 
\bibitem[\protect\citeauthoryear{Patterson}{1994}]{patterson1994} Patterson J., 1994, PASP, 106, 209 
\bibitem[\protect\citeauthoryear{Pereyra}{2000}]{pereyra2000} A. Pereyra, {\em Dust and Magnetic Fields in Dense Regions of the Interstellar Medium} (2000) PhD Thesis, Univ. S\~ao Paulo.
\bibitem[\protect\citeauthoryear{Potter, Hakala \& Cropper}{1998}]{potter1998} Potter S.~B., Hakala P.~J., Cropper M., 1998, MNRAS, 297, 1261 
\bibitem[\protect\citeauthoryear{Press et al.}{1992}]{press1992} Press W.~H., Teukolsky S.~A., Vetterling W.~T., Flannery B.~P., 1992, Numerical Recipes in C, 2nd edn. Cambridge Univ. Press, Cambridge
\bibitem[\protect\citeauthoryear{Ramsay et al.}{1999}]{ramsay1999} Ramsay G., Potter S.~B., Buckley D.~A.~H., Wheatley P.~J., 1999, MNRAS, 306, 809 
\bibitem[\protect\citeauthoryear{Ramsay et al.}{2000}]{ramsay2000} Ramsay G., Potter S., Cropper M., Buckley D.~A.~H., Harrop-Allin M.~K., 2000, MNRAS, 316, 225 
\bibitem[\protect\citeauthoryear{Ramsay et al.}{2004}]{ramsay2004} Ramsay G., Cropper M., Wu K., Mason K.~O., C{\'o}rdova F.~A., Priedhorsky 
W., 2004, MNRAS, 350, 1373 
\bibitem[\protect\citeauthoryear{Raymond}{2009}]{raymond2009} Raymond J.~C., 2009, A\&A, 500, 311
\bibitem[\protect\citeauthoryear{Rodrigues, Cieslinski \& Steiner}{1998}]{rodrigues1998} Rodrigues C.~V., Cieslinski D., Steiner J.~E., 1998, A\&A, 335, 979 
\bibitem[\protect\citeauthoryear{Rodrigues et al.}{2011}]{rodrigues2011} Rodrigues C.~V., Costa J.~E.~R., Silva K.~M.~G., de Souza C.~A., Cieslinski D., Hickel G.~R., 2011, arXiv, arXiv:1101.5554 
\bibitem[\protect\citeauthoryear{Saxton et al.}{2007}]{saxton2007} Saxton C.~J., Wu K., Canalle J.~B.~G., Cropper M., Ramsay G., 2007, MNRAS, 
379, 779 
\bibitem[\protect\citeauthoryear{Silva, Rodrigues \& Costa}{2011a}]{silva2011a} Silva K.~M.~G., Rodrigues C.~V., Costa J.~E.~R., 2011a, xru..conf, 286 
\bibitem[\protect\citeauthoryear{Silva, Rodrigues \& Costa}{2011b}]{silva2011b} Silva K.~M.~G., Rodrigues C.~V., Costa J.~E.~R., 2011b, arXiv, arXiv:1101.5568
\bibitem[\protect\citeauthoryear{Simmons \& Stewart}{1985}]{simmons1985} Simmons J.~F.~L., Stewart B.~G., 1985, A\&A, 142, 100 
\bibitem[\protect\citeauthoryear{Sirk \& Howell}{1998}]{sirk1998} Sirk M.~M., Howell S.~B., 1998, ApJ, 506, 824
\bibitem[\protect\citeauthoryear{Suleimanov et al.}{2008}]{suleimanov2008} Suleimanov V., Poutanen J., Falanga M., Werner K., 2008, A\&A, 491, 525 
\bibitem[\protect\citeauthoryear{Tody}{1986}]{iraf2} Tody D., 1986, SPIE, 627, 733 
\bibitem[\protect\citeauthoryear{Tody}{1993}]{iraf1} Tody D., 1993, ASPC, 52, 173 
\bibitem[\protect\citeauthoryear{Terada et al.}{2001}]{terada2001} Terada Y., Ishida M., Makishima K., Imanari T., Fujimoto R., Matsuzaki K., Kaneda H., 2001, MNRAS, 328, 112 
\bibitem[\protect\citeauthoryear{Thomas \& Reinsch}{1995}]{thomas1995} Thomas H.-C., Reinsch K., 1995, IAUC, 6261, 1 
\bibitem[\protect\citeauthoryear{Thomas \& Reinsch}{1996}]{thomas1996} Thomas H.-C., Reinsch K., 1996, A\&A, 315, L1 
\bibitem[\protect\citeauthoryear{Tovmassian et al.}{2000}]{tovmassian2000} Tovmassian G.~H., et al., 2000, NewAR, 44, 55P 
\bibitem[\protect\citeauthoryear{Vaillancourt}{2006}]{vaillancourt2006} Vaillancourt J.~E., 2006, PASP, 118, 1340
\bibitem[\protect\citeauthoryear{Warren, Sirk \& Vallerga}{1995}]{warren1995} Warren J.~K., Sirk M.~M., Vallerga J.~V., 1995, ApJ, 445, 909
\bibitem[\protect\citeauthoryear{Watson et al.}{1989}]{watson1989} Watson M.~G., King A.~R., Jones M.~H., Motch C., 1989, MNRAS, 237, 299 
\bibitem[\protect\citeauthoryear{Wu, Chanmugam \& Shaviv}{1995}]{wu1995} Wu K., Chanmugam G., Shaviv G., 1995, ApJ, 455, 260 
\end{thebibliography}
\end{document}